\documentclass[pdftex,12pt]{article}
\usepackage{natbib}
\usepackage{graphicx}
\usepackage{multirow}
\usepackage{amsmath, amsthm, amssymb}

\usepackage{hyperref}
\hypersetup{
    bookmarks=true,							
    unicode=true,         					
    pdftoolbar=true,							
    pdfmenubar=false,						
    pdffitwindow=false,						
    pdftitle={Some Advances in Model Selection of Nonparametric and Semiparametric Modeling},					
    pdfauthor={Zaili Fang},									
    pdfsubject={Ph.D. Dissertation Proposal of Statistics},	
    pdfcreator={Zaili Fang},				
    pdfnewwindow=true,						
    colorlinks=false,						
}
\newcommand{\Appendix}{\appendix\def\thesection{Appendix~\Alph{section}}\def\thesubsection{\Alph{section}.\arabic{subsection}}}

\def\aln{\begin{align*}}
\def\eln{\end{align*}}
\def\be{\begin{equation}}
\def\ee{\end{equation}}
\def\bse{\begin{eqnarray*}}
\def\ese{\end{eqnarray*}}
\def\bea{\begin{eqnarray}}
\def\eea{\end{eqnarray}}
\def\cov{\hbox{cov}}

\def\var{\hbox{var}}
\def\var{\hbox{Var}}
\def\cov{\hbox{Cov}}

\def\Tr{\hbox{Tr}}

\def\bdx{\boldsymbol x}
\def\bdz{\boldsymbol z}

\def\mby{\mathbf y}

\def\ANNALS{{\it The Annals of Statistics}}
\def\BMCS{{\it Biometrics}}
\def\BIOK{{\it Biometrika}}
\def\BIOI{{\it Bioinformatics}}

\def\ISR{{\it International Statistical Review}}
\def\JASA{{\it Journal of the American Statistical Association}}

\def\JRSSB{{\it Journal of the Royal Statistical Society, Series B}}
\def\JMLR{{\it Journal of Machine Learning Research}}

\def\TECH{{\it Technometrics}}

\def\cov{\hbox{cov}}

\def\var{\hbox{var}}

\def\wh{\widehat}

\def\frac#1#2{{#1\over#2}}

\def\JASA{{\it Journal of the American Statistical Association}}

\def\wh{\widehat}
\def\ANNALS{{\it Annals of Statistics}}
\def\BIOK{{\it Biometrika}}

\def\JRSSB{{\it Journal of the Royal Statistical Society, Series B}}
\def\BMCS{{\it Biometrics}}

\def\TECH{{\it Technometrics}}

\def\boxit#1{\vbox{\hrule\hbox{\vrule\kern6pt
          \vbox{\kern6pt#1\kern6pt}\kern6pt\vrule}\hrule}}

\def\var{\hbox{Var}}
\def\cov{\hbox{Cov}}

\def\Tr{\hbox{Tr}}

\def\m{\Theta}

\def\ki0x{K_h(X_i-x_0)}
\def\kj0x{K_h(X_j-x_0)}

\def\Li0x{L_{ni}(x_0)}

\def\hpi0x{\psi\{\Y_i,\wh{\m}(x_0)\}}

\def\ci0x{\chi\{\Y_i,\m(x_0)\}}
\def\cj0x{\chi\{\Y_j,\m(x_0)\}}

\def\bi0x{B_n^{-1}(x_0)}

\def\h0b{\wh{\m}(x_0)}

\def\ch0b{\wh{\m}_c(x_0)}

\def\fg0i{\{f_x(x_0)g(x_0)\}^{-1}}

\begin{document}
\font\fmain=cmbx10 scaled\magstep4
\title{Semiparametric Mixed Model for Evaluating Pathway-Environment Interaction}
\author{Zaili Fang$^{1}$, Inyoung Kim$^{1*}$, and Jeesun Jung$^{2}$}
\date{\today}
\maketitle
\thispagestyle{empty}
\baselineskip=26pt
\hskip 5mm \\
1 Department of Statistics, Virginia Polytechnic
  Institute and State University, Blacksburg, Virginia, U.S.A.\\
2 Department of Medical and Molecular Genetics, Indiana University School of Medicine, Indianapolis, U.S.A.
\hskip 5mm \\
\hskip 5mm \\
\noindent
*To whom correspondence should be addressed:\\
Inyoung Kim, Ph.D.\\
Department of Statistics, Virginia Polytechnic
  Institute and State University, 410A Hutcheson Hall, Blacksburg, VA 24061-0439, U.S.A.\\
Tel: (540) 231-5366\\
Fax: (540) 231-3863\\
Email: inyoungk$@$vt.edu\\
\hskip 5mm \\
\newpage
\begin{abstract}
A biological pathway represents a set of genes that serves a particular cellular or a physiological function. The genes within the same pathway are expected to function together and hence may interact with each other.  It is also known that many genes, and so pathways, interact with other environmental variables. However, no formal procedure has yet been developed to evaluate the pathway-environment interaction. In this article, we propose a semiparametric method to model the pathway-environment interaction. The method connects a least square kernel machine and  a semiparametric mixed effects model. We model nonparametrically the environmental effect via a natural cubic spline. Both a pathway effect and an interaction between a pathway and an environmental effect are modeled nonparametrically via a kernel machine, and we estimate variance component representing an interaction effect under a semiparametric mixed effects model. We then employ a restricted likelihood ratio test and a score test to evaluate the main pathway effect and the pathway-environment interaction. The approach was applied to a genetic pathway data of Type II diabetes, and pathways with either a significant main pathway effect, an interaction effect or both were identified. Other methods previously developed determined many as having a significant main pathway effect only. Furthermore, among those significant pathways, we discovered some pathways having a significant pathway-environment interaction effect, a result that other methods would not be able to detect.

\vskip 5mm
\noindent
\underline{\hbox{\bf Keywords}}: environmental variable; Gaussian random process; Kernel machine; Pathway analysis; Semiparametric model; Smoothing splines
\vskip 5mm
\noindent
\underline{\hbox{\bf Running Title}}: Semiparametric Method for Evaluating Pathway-Environment Interaction
\thispagestyle{empty}
\end{abstract}
\newpage
\section{Introduction}\label{a_sec1}
Gene-related diseases are complex processes associated not only with specific gene or gene sets but also with gene-gene and gene-environment interaction. For decades, statistical methods have focused on analyzing microarray data based on single genes or single-nucleotide polymorphisms (SNPs) analysis \citep{a5,a19,a29,a32,a40}. However, single-gene based methods have many limitations. For instance, the effect of one gene on a disease is difficult to interpret and current methods are unable to model gene dependencies so that they may not detect genes with moderate changes that give more insight into biological processes but pick up single gene with dramatic changes \citep{b21}. For these reasons, gene-set or pathway-based approaches have attracted increasing attention in recent years \citep{a10,a11,b17,a45,a35,a37,a22}. It is recognized that a joint study of the association between the outcome and a group of genes within the same pathway could complement genes/SNPs analysis for providing insight in understanding complex diseases \citep{a45}.

A genetic pathway is the interactions of genes that depend on each other's individual functions and act accordingly to create the aggregate function related to a cellular process \citep{a10}. There are several special characteristics of pathways, such as various dimensionality (a pathway can contain several genes or over a thousand ones), and interaction network (genes within the a pathway are expected to function together and hence interact with each other). Thus traditional statistical analyses face difficulties in handling these situations. For instance, linear parametric models usually either fail due to the ``curse of dimensionality'', or end up with computational explosion in the number of possible interactions among genes within a pathway. To deal with these difficulties, many innovative statistical methods have merged in recent years. \citet{a10} proposed a global test derived from a random effects model to determine the significance of the global expression pattern of a group of genes. A random forests approach was proposed by \citet{a35}. \citet{b17} proposed a semiparametric model for covariate and genetic pathway effects on continuous outcomes, where the covariate effects and the pathway effect are modeled parametrically and nonparametrically, respectively. They established the connection between the least squares kernel machine (LSKM) and linear mixed models, which simplifies specification of a nonparametric model with multi-dimensional data. \citet{a37} considered more complicated situations with two or more pathway effects presented in the linear mixed model, which allows the researcher to study how multiple pathways relate to the phenotype of interest. A semiparametric Bayesian approach has also been proposed for evaluating pathway effects on clinical outcomes \citet{a22}. However, despite the success of analyzing pathways instead of a single gene, all existing methods ignore the environment exposure covariates, and still fewer focus on the interaction between environmental variables and the genetic pathways.

It has been recognized that genetic factors alone cannot account for many cases of gene related disease \citep{a1,a4}. The gene-environment (G-E) or pathway-environment (P-E) interactions are critical in understanding the dynamic process of disease since ignoring them may mask the detection of a genetic effect and may lead to inconsistent association results \citep{a30}. Furthermore, understanding the G-E interactions can be important for risk prediction and evaluating the benefit of changes in modifiable environmental exposures or environmental regulations. For these reasons, the number of studies utilizing gene-environment interactions has increased dramatically. These range from semiparametric linear or logistic regression models with linear combinations of genes/SNPs as the predictor \citep{a5,a29,a38} to the multifactor dimensionality reduction (MDR) as a data mining technique for identifying genetic and environmental effects associated with either dichotomous or continuous phenotypes \citep{a40,a19,a32}. Unfortunately, these studies are all genes/SNPs based methods, and they possess problems in dealing with the pathway analysis. For example, representing the pathway effects with linear combinations of genes has limitations in detecting non-linear patterns of interacting genes. Furthermore, the number of genes in a pathway can be in the hundreds or thousands, which makes modeling the gene-gene or gene-environment interaction very consuming. 

To capture high order interactions within the high dimensional genes regressor space as well as the G-E interactions, \citet{b35} employed a nonparametric regression model with a Gaussian process. With their model the gene and environmental variables are modeled non-parametrically, and all of the possible interactions effects are considered simultaneously. However, using one Gaussian process to describe both gene and environmental variable function spaces results in all the interaction effect being indistinguishable. Thus it is almost impossible to apply a suitable test for interesting effects such as G-E interaction.

In this paper, we propose a semiparametric mixed effects model to include environmental variables, genetic pathway effect, and their interaction. By extending \citet{b17}'s linear mixed model to our model, we evaluate the interaction between an environmental variable and pathway as well as allow nonlinear relationships between the environmental variable and a continuous outcome. Assuming that both the pathway and interaction effects have multivariate normal distributions with a zero mean and covariance structure with specific kernels, we model them within the framework of Gaussian processes. Thus in our model both pathway and interaction effects are indeed modeled as random effects. Instead of modeling the environmental variable as a parametric fixed effect, we model it non-parametrically via natural cubic spline. By modeling environmental variables and pathways in this way, we can construct the kernel for the P-E interaction based on the analysis-of-variance-like (ANOVA-like) decompositions of functions \citep{a46,a15} for a multivariate function. 
The feature of our method is to model the interaction between environmental and pathway covariates separately from the interactions among genes within the pathway, which are automatically modeled by the Gaussian process for pathway effect. Our model also extends the additive and interaction smoothing splines for univariate functions to multivariate functions with arbitrary kernel.

In a mixed model, the smoothing parameters of the spline and the Gaussian kernels can be considered as the variance components of the random effects, and thus are simultaneously estimated by maximizing the restricted maximum likelihood (REML). By additively modeling the multivariate functions, this model is suitable for analyzing genetic pathway data in which the P-E interaction attracts particular interests. Furthermore, the covariance structure of our model makes the test of the ``overall'' pathway effect or P-E interaction effect possible. By ``overall'' we mean either the main effect of a pathway, the interaction effect associated with the pathway, or both. The restricted likelihood ratio test (RLRT) of two zero variance components under non-standard conditions is employed to test the overall pathway effect, while the RLRT of one zero-variance component and score test are applied to test the P-E interaction.

We first define our model in Section \ref{a_sec2}, and discuss two REML methods to estimate the model parameters in Section \ref{a_sec3}. Then in Section \ref{a_sec4}, we introduce PLRT statistics for testing two or one zero-variance components and the score test for testing the P-E interaction effect. In Section \ref{a_sec5}, we present a set of simulation studies concerning nonparametric function estimates and variance component tests for various settings. In Section \ref{a_sec6} we apply our method to the  genetic pathway data for Type II diabetes. Finally, in the last Section, we conclude our work and discuss potential extensions of our model. 

\section{Construction of Semiparametric Linear Mixed Effects Models}\label{a_sec2}
\subsection{Model Description and the Kernel of the Interaction Function Space}\label{a_sec2_1}
Let us consider that we have a total of $n$ subjects and the $i$th subject has a continuous disease-related outcome $y_i, i=1,2,...n$. We are interested in relating this response $\mby=(y_1,y_2,...,y_n)^T$ with one particular pathway gene expression data $Z=({\bdz}_1,{\bdz}_2,...,{\bdz}_n)^T$ and $k$ environmental variables. In a general form, we can write this nonlinear relationship as
\be\label{a_e1}
\mby=\mathbf{f}+\boldsymbol{\epsilon},
\ee
where $\boldsymbol{\epsilon}$ and $\mathbf{f}$ are $n\times1$ dimensional vectors with a specific relationship with $\mathbf y$ for the $i$th entry as $y_i=f({\bdx}^T_i,{\bdz}^T_i)+\epsilon_i$, in which ${\bdx}^T_i=(x_{i1},x_{i2},...,x_{ik})$ is $1\times k$ vector of environmental variables and ${\bdz}^T_i=(z_{i1},z_{i2},...,z_{ip})$ is the $1\times p$ vector of gene expression within a pathway and $p$ is the gene number. In this paper, we only consider the case with one environmental variable, i.e., $k=1$ so that the input $\bdx^T$ is reduced to univariate $x$. We assume that the errors $\boldsymbol{\epsilon}\sim N(0,\sigma^2I)$ are $n\times1$ iid random variables vector. $f(\cdot)$ denotes the unknown non-linear smooth functions for $x_i$, ${\bdz}^T_i$, and their interaction. In this paper, we assume function $f$ has the following form:
\be\label{a_e2}
f(x,{\bdz}^T)=\beta_0+f_x(x)+f_z({\bdz}^T)+f_{xz}(x,{\bdz}^T),
\ee
where $\beta_0$ is the intercept term, and $f_\alpha, \alpha\in\{x, z, xz\}$, represents the nonlinear effect of the environmental variable, the pathway or the interaction respectively. The above equation is similar to the additive model with two univariate variables and their interaction, except ${\bdz}^T$ is a multivariate variable. By writing the general model (\ref{a_e1}) in this way, we can estimate $f_x$, $f_z$ and their interaction $f_{xz}$ separately according to the characteristics of the pathway and the environmental variable.  We model $f_x(x)$ using the nonparametric function such as a cubic smoothing spline \citep{a46,a25,a48}. To handle the high dimensional pathway covariates, ${\bdz}^T$, we may consider a Gaussian process to express $f_z({\bdz}^T)$ since the least squares kernel machine method with the Gaussian kernel has achieved success in a genetic pathway data analysis \citep{b17}.

Before we derive the specific representation for the interaction function, we need examine the function space of $f_x$ and $f_z$ respectively. For the smoothing spline $x\in\mathcal{T}=[0,1]$, $f_x$ is spanned on the function space $\mathcal{H}_x=\mathcal{H}_x^0\bigoplus\mathcal{H}_x^1$, where ``$\bigoplus$'', $\mathcal{H}_x^0$ and $\mathcal{H}_x^1$ represent the direct sum operator of two subspaces, the null function space and the penalized function space respectively \citep{a46}. Assuming $n$ distinct values of $x_i$ such that $0<x^0_1<\cdots<x^0_n<1$, the $m$th order smoothing spline estimator $f_x(x)$ can be expressed as \citep{a46,a48},
\[\label{a_e5}
f_x(x)=\sum_{j=1}^mb_j\phi_j(x)+\sum_{i=1}^nc_ik_x(x,x^0_i),
\]
where $\phi_j(x)$ is the polynomial basis that span the null space $\mathcal{H}_x^0$ with $\phi_j(x)=x^{j-1}/(j-1)!, j=1,2,\dots,m$, and $k_x(x,x^0_i)=[(m-1)!]^{-2}\int(x-u)^{m-1}_+(x^0_i-u)_+^{m-1}du$ is the kernel which uniquely determines the space $\mathcal{H}_x^1$.
For $m=2$, the natural cubic spline that we shall apply in our model, the kernel of $\mathcal{H}_x^1$ can be calculated as \citep{a18,b25}
\be\label{a_e7}
k_x(x,x')=\int_0^1(x-u)_+(x'-u)_+du={{\min(x,x')^3}\over3}+{{\min(x,x')^2|x-x'|}\over2},
\ee
where subscript ``+'' indicates the positive part of the expression. For the null space $\mathcal{H}_x^0$, the kernel is calculated as $k^0_x(x,x')=\sum_{j=1}^2\phi_j(x)\phi_j(x')=1+xx'$.

With the orthonormal polynomial basis, $\mathcal{H}_x^0=\{1\}\bigoplus\{x\}$, where $\{1\}$ and $\{x\}$ stand for the linear function spaces spanned by the constant 1 and the linear basis $x$ which is centered \citep{a15a}. Since the kernel of the function space of the direct sum of two subspaces is expressed by the direct sum of the kernel of the subspaces \citep{a0,a46}, we can derive the kernel of the function space without the the constant term for the cubic smoothing spline, $\{x\}\bigoplus\mathcal{H}_x^1$, as $[xx'+k_x(x,x')]$.

For the function space of $f_z$, we consider a similar argument by \citet{a28} that starting from a parametric model, we can span the function of $f_z$ by a radial basis \be\label{a_e7a}
f_z({\bdz}^T)=\sum_{h=1}^Hc_h\phi_h({\bdz}^T),
\ee
where $\phi_h({\bdz}^T)=\exp\left[-{\|{\bdz}-{\bdz}_h\|^2}\over{2\rho}\right]$ is the radial basis functions centered at fixed points $\{{\bdz}_h\}_{h=1}^H$. Assuming $\mathbf c=(c_1,...,c_h)^T\sim N(0, \tau_zI)$, the entry of the covariance matrix of $\mathbf f_z$ is expressed as
\[
R=\tau_z\sum_h\phi_h({\bdz})\phi_h({\bdz}').
\]

Taking as an example a one-dimensional case, \citet{a28} shows that in the above expression the sum over $h$ becomes an integral when taking the limit $H\rightarrow\infty$ such that $R=\tau_z\exp\left[-(z-z')^2/\rho\right]$. Generalizing from this particular case, we can define the Gaussian kernel of the function space $\mathcal{H}_z^1$ on ${\bdz}$
\be\label{a_e8}
k_z\left({\bdz}^T,{{\bdz}'}^T\right)=\exp\left(-\|{\bdz}-{\bdz}'\|^2/\rho\right),
\ee
and we assume that $f_z$ is generated from a zero mean Gaussian process with the kernel matrix produced by $k_z$.

Since the tensor product of the kernels of two function spaces determines a new function space \citep{a0}, we use the tensor product of the kernels of $\{x\}\bigoplus\mathcal{H}_x^1$ and $\mathcal{H}_z^1$ to construct a new function space, $\mathcal{H}_{xz}^1$, which contains any order interaction $f_{xz}$ between $x$ and ${\bdz}^T$. Now we can express the kernel of the interaction function space as
\be\label{a_e9}
k_{xz}\left(x,{\bdz}^T;x',{{\bdz}'}^T\right)=\left[xx'+k_x(x,x')\right]\cdot k_z\left({\bdz}^T,{{\bdz}'}^T\right).
\ee
Therefore, we are able to represent the nonparametric interaction function using a zero mean Gaussian process with the kernel matrix produced by this kernel function.

In the rest of this paper, we use $K_x$, $K_z$ and $K_{xz}$ to stand for the Gram or kernel matrices produced by $k_x, k_z$ and $k_{xz}$ respectively. In a specific problem, the environmental variable $x$ must be scaled into $\mathcal{T}=[0,1]$ to construct the interaction kernel. Notice the model expression (\ref{a_e2}) is not the analysis of variance (ANOVA) decomposition of the smoothing function $f$ since $\mathcal{H}_z^1$ and $\mathcal{H}_{xz}^1$ are not orthogonal to each other. This may cause the identifiability problem between $f_z$ and $f_{xz}$. However, in practice, this problem only happens to our model in extreme situations such as when the entries of matrix $xx'+k_x(x,x')$ are close to each other. In general, $f_z$ and $f_{xz}$ can be identified well as shown in the simulation and application study.

\subsection{Linear Mixed model Representation}\label{a_sec2_3}
Now we are prepared to pose the optimization problem. Based on the above argument, the corresponding function spaces that are penalized are $\mathcal{H}_x^1$, $\mathcal{H}_z^1$ and $\mathcal{H}_{xz}^1$. Analogous to the additive models \citep{b11}, the estimation problem for model (\ref{a_e1}) becomes: for a given set of predictors $(x_i,\bdz_i^T), i=1,2,...,n$, find $f$ to maximize
\be\label{a_e10}
-{1\over2}(\mathbf{y}-\mathbf{f})^T(\mathbf{y}-\mathbf{f})-{1\over2}\sum_\alpha\lambda_\alpha\|f_\alpha\|^2_{\mathcal{H}_\alpha^1},
\ee
where $\|f_\alpha\|_{\mathcal{H}^1_\alpha}$'s are the norms induced by $K_\alpha$ of $\mathcal{H}_\alpha^1, \alpha\in\{x, z, xz\}$, and $\lambda_\alpha$'s are the penalty parameters that balance the tradeoff between goodness-of-fit and smoothing of the curve or high dimensional surface. The solutions to expression (\ref{a_e10}) are called the least square kernel machine estimation, and \citet{b17} showed the equivalence of the least square kernel machine to the linear mixed model without interaction effects.

The model (\ref{a_e2}) can be represented in terms of a mixed model as follows. According to the Representer Theorem \citep{a23}, the nonparametric function can be expressed by the kernel, $f_z(\cdot)=\sum_{i=1}^n a_ik_z(\cdot, \bdz^T_i)$ and $f_{xz}(\cdot)=\sum_{i=1}^n b_ik_{xz}(\cdot; x_{i}, {\bdz}^T_i)$. So the vectors of these functions are
\bse\label{a_e11}
\begin{array}{ll}
\mathbf f_z&=K_z\mathbf a,\\
\mathbf f_{xz}&=K_{xz}\mathbf b,
\end{array}
\ese
where $a_i\in\mathbb{R}, b_i\in\mathbb{R}$. Based on the properties of reproducing kernels, the squared norms of $\mathcal{H}_z^1$ and $\mathcal{H}_{xz}^1$ can be expressed as
\bse\label{a_e12}
\begin{array}{ll}
\|f_z\|^2_{\mathcal{H}_z^1}&=\mathbf a^TK_z\mathbf a=\mathbf f_z^TK_z^{-1}\mathbf f_z,\\
\|f_{xz}\|^2_{\mathcal{H}_{xz}^1}&=\mathbf b^TK_{xz}\mathbf b=\mathbf f_{xz}^TK_{xz}^{-1}\mathbf f_{xz}.
\end{array}
\ese

To represent the remaining part of model (\ref{a_e2}), $\beta_0+f_x(\cdot)$, we follow \citet{a25,a47,a13,b9}'s procedure. The vector of $f_x$, $\mathbf{f}_x$ (note here the constant $\beta_0$ is absorbed into $f_x$), can be expressed in terms of $\boldsymbol\beta=(\beta_0, \beta_1)^T$ and $(n-2)\times1$ random vector $\mathbf r_x$ as
\be\label{a_e13}
\mathbf{f}_x=X\boldsymbol{\beta}+B\mathbf r_x
\ee
for $n$ distinct input $x$ values, where $\mathbf r_x\sim N(0, \tau_xI)$ and $X$ is the design matrix of the null space $\mathcal{H}_x^0$ spanned by the orthogonal polynomial basis, i.e., $X=(\mathbf{1}, \mathbf{x})$ and $\mathbf{x}$ is the $n\times1$ vector of centered $x$. $B$ is a matrix defined as $B=L(L^TL)^{-1}$, where $L$ is $n\times(n-2)$ full rank matrix with $M=LL^T$. $M$ is a penalty matrix defined by \citet{b9} such that the squared norm of $\mathcal{H}_x^1$,
\[\label{a_e14}
\|f_x\|^2_{\mathcal{H}_x^1}=\int_0^1\left[f_x''(t)\right]^2dt=\mathbf f_x^TM\mathbf f_x=\mathbf r_x^T\mathbf r_x.
\]
More details to define $B$ and $M$ can be found in \citet{b9}, \citet{a47} and \ref{ap1}.

Plugging those representations of square norms and $\mathbf f_\alpha$'s back into (\ref{a_e10}), we have
\[\label{a_e15}
-{1\over2}(\mathbf{y}-\mathbf{f})^T(\mathbf{y}-\mathbf{f})-{1\over2}\left(\lambda_x\mathbf r_x^T\mathbf r_x+\lambda_z\mathbf f_z^TK_z^{-1}\mathbf f_z+\lambda_{xz}\mathbf f_{xz}^TK_{xz}^{-1}\mathbf f_{xz}\right).
\]
If we define $\lambda_x=\sigma^2/\tau_x, \lambda_z=\sigma^2/\tau_z$ and $\lambda_{xz}=\sigma^2/\tau_{xz}$, and have random vectors $\mathbf r_z=\mathbf f_z, \mathbf r_z\sim N(0,\tau_zK_z)$ and $\mathbf r_{xz}=\mathbf f_{xz}, \mathbf r_{xz}\sim N(0,\tau_{xz}K_{xz})$, then the above equation is equivalent to
\be\label{a_e16}
-{1\over{2\sigma^2}}(\mathbf{y}-\mathbf{f})^T(\mathbf{y}-\mathbf{f})-{1\over{2\tau_{x}}}\mathbf r_x^T\mathbf r_x-{1\over{2\tau_{z}}}\mathbf{r}_z^TK_z^{-1}\mathbf{r}_z-{1\over{2\tau_{xz}}}\mathbf{r}_{xz}^TK_{xz}^{-1}\mathbf{r}_{xz},
\ee
which is the triple penalized  log likelihood function of the linear mixed model
\be\label{a_e17}
\mathbf{y}=\mathbf{f}+\boldsymbol{\epsilon}=X\boldsymbol{\beta}+B\mathbf r_x+\mathbf{r}_z+\mathbf{r}_{xz}+\boldsymbol{\epsilon}.
\ee
From the Bayesian point-of-view, $\mathbf f$ is interpreted as the sum of four zero-mean stationary Gaussian processes, each with a prior covariance function $\tau_\alpha K_\alpha$ ($\boldsymbol\beta$ can be viewed with infinite variance). The vectors $\mathbf r_z$ and $\mathbf r_{xz}$ have more specific meanings as the pathway main effect and the P-E interaction effect. Although $\mathbf r_x$ does not have such a meaning, it can be interpreted as the nonlinear contribution of the relationship of the response and the environmental variable.

Differentiating expression (\ref{a_e17}) with respect to $\boldsymbol{\beta}$ and $\mathbf{r}_\alpha$'s, it is easy to show that the best linear unbiased prediction (BLUP) estimate of the random effects, given $\sigma^2$ and $\tau_\alpha$'s as fixed, is obtained from solving
\be\label{a_e18}
\begin{bmatrix} X^TX & X^TB            & X^T                       & X^T \\
                B^TX & B^TB+\lambda_xI & B^T                       & B^T \\
                   X & B               & I+\lambda_z{[K_z]}^{-1} & I   \\
                   X & B               & I                         & I+\lambda_{xz}{[K_{xz}]}^{-1} \end{bmatrix}
\times \left[ \begin{array}{c} \boldsymbol{\beta} \\ \mathbf r_x  \\ \mathbf{r}_z \\ \mathbf{r}_{xz} \end{array} \right]
=      \left[ \begin{array}{c} X^T\mathbf{y}      \\ B^T\mathbf{y} \\ \mathbf{y}   \\ \mathbf{y}      \end{array} \right].
\ee
Equation (\ref{a_e18}) shows that the BLUP estimate of $\boldsymbol{\beta}$ and $\mathbf{r}_\alpha$'s are unique if $X^TX$ is full rank which is usually satisfied.
\subsection{Estimate Pathway and Interaction Effects}\label{a_sec2_4}
Given the fixed parameters $\sigma^2$ and $\tau_\alpha$'s, the covariance of $\mathbf{y}$ is obtained as follows using model (\ref{a_e17}),
\be\label{a_e19}
\Sigma=\cov(\mathbf{y})=\sigma^2I+\tau_xBB^T+\tau_zK_z+\tau_{xz}K_{xz}.
\ee
Instead of solving expression (\ref{a_e18}) directly, we perform recursive steps to simultaneously achieve the approximate expressions of $\boldsymbol{\beta}$ and $\mathbf{r}_\alpha$'s, $\alpha\in \{x,z,xz\}$,
\be\label{a_e20}
\begin{array}{ll}
\hat{\boldsymbol{\beta}}&=\left(X^T\Sigma^{-1}X\right)^{-1}X^T\Sigma^{-1}\mathbf y,\\
\hat{\mathbf r}_x&=\left(B^T\Delta_1^{-1}B+\tau_x^{-1}I\right)^{-1}B^T\Delta_1^{-1}(\mathbf y-X\hat{\boldsymbol{\beta}}),\\
\hat{\mathbf r}_z&=\left(\Delta_2^{-1}+\tau_z^{-1}K_z^{-1}\right)^{-1}\Delta_2^{-1}(\mathbf y-X\hat{\boldsymbol{\beta}}-B\hat{\mathbf r}_z),\\
\hat{\mathbf r}_{xz}&=\left(\Delta_3^{-1}+\tau_{xz}^{-1}K_{xz}^{-1}\right)^{-1}\Delta_3^{-1}(\mathbf y-X\hat{\boldsymbol{\beta}}-B\hat{\mathbf r}_x-\hat{\mathbf r}_z),
\end{array}
\ee
where $I$ is the $(n-2)\times(n-2)$ identity matrix, and $\Delta_j,\;j=1,2,3$, are covariances for the following distributions,
\be\label{a_e21}
\begin{array}{ll}
\mathbf y=X\boldsymbol{\beta}+\boldsymbol{\epsilon}_0, &\boldsymbol{\epsilon}_0\sim N(0,\Delta_0=\Sigma),\\
\mathbf y-X\hat{\boldsymbol{\beta}}=B\mathbf r_x+\boldsymbol{\epsilon}_1, &\boldsymbol{\epsilon}_1\sim N(0,\Delta_1=\sigma^2I+\tau_zK_z+\tau_{xz}K_{xz}),\\
\mathbf y-X\hat{\boldsymbol{\beta}}-B\hat{\mathbf r}_x=\mathbf r_z+\boldsymbol{\epsilon}_2, &\boldsymbol{\epsilon}_2\sim N(0,\Delta_2=\sigma^2I+\tau_{xz}K_{xz}),\\
\mathbf y-X\hat{\boldsymbol{\beta}}-B\hat{\mathbf r}_x-\hat{\mathbf r}_z=\mathbf r_{xz}+\boldsymbol\epsilon, &\boldsymbol\epsilon\sim N(0,\Delta_3=\sigma^2I).
\end{array}
\ee
The above expressions for $\hat{\boldsymbol{\beta}}$ and $\hat{\mathbf{r}}_\alpha$'s are all linear transformations of $\mathbf{y}$; thus, their covariances are easily determined using identity $\cov(A\mathbf{y})=A\cov(\mathbf{y})A^T=A\Sigma A^T$, where $A$ is the transformation matrix in expressions (\ref{a_e20}).
\section{REML Estimation of the Variance Components}\label{a_sec3}
\subsection{REML Approach for Estimating Variance Components}\label{a_sec3_1}
In the previous Section, when solving the equation (\ref{a_e18}) we assume that the regularization parameters, $\tau_x$, $\tau_z$ and $\tau_{xz}$, the scale parameter $\rho$ for Gaussian processes, and the error variance $\sigma^2$ are already known. In this linear mixed model framework, we can estimate the parameter $\boldsymbol{\theta}=(\sigma^2, \tau_x,\tau_z,\tau_{xz}, \rho)^T$ simultaneously using restricted maximum likelihood (REML) estimation. REML is superior to the maximum likelihood (ML) method in terms of adjusting the small sample bias \citep{a48}. The REML of our model is derived routinely \citep{a16} up to the usual additive constant
\be\label{a_e22}
l_{R}=-{1\over2}\log|\Sigma|-{1\over2}|X^T\Sigma^{-1}X|-{1\over2}(\mathbf{y}-X\hat{\boldsymbol{\beta}})^T\Sigma^{-1}(\mathbf{y}-X\hat{\boldsymbol{\beta}})+c,
\ee
where $c$ is constant. Another advantage of using REML is that it accounts for the degrees-of-freedom adjustment of replacing $\boldsymbol{\beta}$ with $\hat{\boldsymbol{\beta}}$ in expression (\ref{a_e22}) \citep{a3}. Taking the derivatives of (\ref{a_e22}) with respect to $\boldsymbol{\theta}$, the estimates of $\boldsymbol{\theta}$ are obtained by solving
\be\label{a_e23}
\begin{split}
{{\partial l_{R}}\over{\partial\sigma^2}}&=-{1\over2}\Tr(P)+{1\over2}(\mathbf{y}-X\hat{\boldsymbol{\beta}})^T\Sigma^{-1}\Sigma^{-1}(\mathbf{y}-X\hat{\boldsymbol{\beta}})=0,\\
{{\partial l_{R}}\over{\partial\tau_\alpha}}&=-{1\over2}\Tr\left({{\partial\Sigma}\over{\partial \tau_\alpha}}P\right)+{1\over2}(\mathbf{y}-X\hat{\boldsymbol{\beta}})^T\Sigma^{-1}{{\partial \Sigma}\over{\partial \tau_\alpha}}\Sigma^{-1}(\mathbf{y}-X\hat{\boldsymbol{\beta}})=0,\;\; \alpha\in\{x, z, xz\},\\
{{\partial l_{R}}\over{\partial \rho}}&=-{1\over2}\Tr\left({{\partial \Sigma}\over{\partial \rho}}P\right)+{1\over2}(\mathbf{y}-X\hat{\boldsymbol{\beta}})^T\Sigma^{-1}{{\partial \Sigma}\over{\partial \rho}}\Sigma^{-1}(\mathbf{y}-X\hat{\boldsymbol{\beta}})=0,
\end{split}
\ee
where $P=\Sigma^{-1}-\Sigma^{-1}X(X^T\Sigma^{-1}X)^{-1}X^T\Sigma^{-1}$, and ${{\partial \Sigma}\over{\partial \rho}}=\tau_z{{\partial K_z}\over{\partial \rho}}+\tau_{xz}{{\partial K_{xz}}\over{\partial \rho}}$. The $5\times5$ information matrix $\mathcal{I}(\boldsymbol{\theta})$ has the $i, j$th entry as
\be\label{a_e24}
\mathcal{I}(\boldsymbol{\theta})_{ij}={1\over2}\Tr\left(P{{\partial \Sigma}\over{\partial \theta_i}}P{{\partial \Sigma}\over{\partial \theta_j}}\right),
\ee
and the variance of $\hat{\boldsymbol\theta}$ can be estimated through the expression of the information matrix. Equation (\ref{a_e23}) can be solved using an iteration method such as Fisher's scoring method. 
In practice, the sample size $n$ may be small, for instance the Type II diabetes data contains only 35 observations, while the model (\ref{a_e17}) includes two fixed-effect parameters and three smoothing parameters. We may have problems with overparameterization, and it may cause a negative estimate of the variance components based on REML. In such case, the step-halving method can be adopted \citep{a20}, but still the corresponding variance component can be estimated as very close to zero. 
\subsection{Profile REML Approach for Estimating Variance Components}\label{a_sec3_2}
In this Section, we suggest a modification to the REML estimation of the variance components so that the estimate of the error components always remains in the parameter space. This new approach makes the use of the profile restricted maximum likelihood (p-REML). The covariance of $\mathbf{y}$ in expression (\ref{a_e19}) can be written as $\Sigma=\sigma^2\Sigma_\lambda$, where $\Sigma_\lambda=(I+\lambda_x^{-1}BB^T+\lambda_z^{-1}K_z+\lambda_{xz}^{-1}K_{xz} )$. Defining the matrix $P_\lambda=\Sigma_\lambda^{-1}-\Sigma_\lambda^{-1}X(X^T\Sigma_\lambda^{-1}X)^{-1}X\Sigma_\lambda^{-1}$, and $P=P_\lambda/\sigma^2$, the restricted log likelihood function (\ref{a_e22}) can be rewritten as
\be\label{a_e26}
l_R=-{1\over2}(n-q)\log(\sigma^2)-{1\over2}|\Sigma_\lambda|-{1\over2}\log|X^T\Sigma_\lambda^{-1}X|-{1\over2}{{\mathbf{y}^TP_\lambda\mathbf{y}}\over\sigma^2}+c,
\ee
where $q=2$ is the rank of $X$. Assuming that $\lambda_\alpha, \alpha\in\{x,z,xz\}$ are known, by solving the derivative of (\ref{a_e26}) with respect to $\sigma^2$ set equal to zero, the p-REML estimator of $\sigma^2$ is
\be\label{a_e27}
\hat{\sigma}^2={{\mathbf{y}^TP_\lambda\mathbf{y}}\over{n-q}}.
\ee
Since $P_\lambda\Sigma_\lambda$ is idempotent, ${{\mathbf{y}^TP_\lambda\mathbf{y}}\over\sigma^2}\sim\chi^2_{\hbox{r}(P_\lambda)}$, where $\hbox{r}(P_\lambda)=\Tr(P_\lambda)$ is the rank of $P_\lambda$, the variance of $\var(\hat{\sigma}^2)\approx2\hat{\sigma}^4\Tr(P_\lambda)/(n-q)^2$. Plug $\hat{\sigma}^2$ back into expression (\ref{a_e26}) and we have the log profile restricted likelihood (PRL) function
\be\label{a_e28}
l_{PR}=-{1\over2}\log|\Sigma_\lambda|-{1\over2}|X^T\Sigma_\lambda^{-1}X|-{{n-q}\over2}\log({\mathbf{y}^TP_\lambda\mathbf{y}})+c.
\ee
Now we can use the similar scoring algorithm to estimate $\boldsymbol{\theta}^*=(\lambda_x^{-1}, \lambda_z^{-1}, \lambda_{xz}^{-1}, \rho)$. By simple algebra the score of the p-REML likelihood is
\be\label{a_e29}
{{\partial l_{PR}}\over{\partial \theta^*_j}}=-{1\over2}\Tr\left({{{\partial\Sigma_\lambda}\over{\partial\theta^*_j}}P_\lambda}\right)+{1\over{2\hat{\sigma}^2}}\mathbf{y}^TP_\lambda{{\partial\Sigma_\lambda}\over{\partial\theta^*_j}}P_\lambda\mathbf{y}, j=1, 2, 3, 4,
\ee
and the $i, j$th entry of the information matrix $\mathcal{I}^*(\boldsymbol{\theta}^*)$ for the PRL can be approximated as
\be\label{a_e30}
\mathcal{I}^*(\boldsymbol{\theta}^*)_{ij}={1\over{2(n-q)}}\Biggl\{(n-q-2)\Tr\left({{\partial \Sigma_\lambda}\over{\partial \theta^*_i}}P_\lambda{{\partial \Sigma_\lambda}\over{\partial\theta^*_j}}P_\lambda\right)-\Tr\left({{\partial\Sigma_\lambda}\over{\partial\theta^*_i}}P_\lambda\right)\Tr\left({{\partial\Sigma_\lambda}\over{\partial\theta^*_j}}P_\lambda\right)\Biggr\}.
\ee
Note that $\mathcal{I}^*(\boldsymbol{\theta}^*)$ is positive definite when $n$ is large enough. \citet{a6} also showed the convergence of $\mathcal{I}^*(\boldsymbol{\theta}^*)$ under regular conditions so that we can apply the restricted likelihood ratio test (RLRT, see Section \ref{a_sec4}). Since PRL is not a true likelihood, we only use PRL for statistical test purposes, and use p-REML to obtain a better estimate of the variance components. The variances of $\boldsymbol{\theta}$ is found by plugging the p-REML estimates into (\ref{a_e24}).
\section{Test for Pathway Effects}\label{a_sec4}
\subsection{Test for Two Zero Variance Components}\label{a_sec4_1}
One of the primary problems in the study of pathway based analysis is testing the ``overall'' pathway effects. Recall that the meaning of ``overall'' refers to either the main effect of a pathway, the interaction effect associated with the pathway, or both. In model (\ref{a_e17}), two random effects are involved with the overall pathway effects. Thus, the hypothesis for testing the overall pathway effect is
\be\label{a_e31}
H_0: \tau_z=\tau_{xz}=0\;\; \textrm{vs.}\;\; H_a: \tau_z>0\; \textrm{or}\; \tau_{xz}>0,
\ee
which is equivalent to the following test
\be\label{a_e32}
H_0: \lambda^{-1}_z=\lambda^{-1}_{xz}=0\;\; \textrm{vs.}\;\; H_a: \lambda^{-1}_z>0\; \textrm{or}\; \lambda^{-1}_{xz}>0.
\ee
For this type of test problem, a likelihood ratio test (LRT) is most commonly used. Note that parameter space for $\boldsymbol{\theta}=(\lambda^{-1}_x,\lambda^{-1}_{z}, \lambda^{-1}_{xz}, \rho)^T$ equals $[0, \infty)^3\times(0,\infty)$ (to avoid abuse of notation, in this Section, $\boldsymbol{\theta}$ and $\mathcal{I}$ stand for counterparts of PRL). The true parameters $\boldsymbol{\theta}_0$ are either in the interior or on the boundary of the parameter space, so the LRT is nonstandard. 
\citet{a44} generalized the hypothesis test for both interior and boundary problems within a setting of mixed regression fitting, so it allows the nonidentically distributed response variable $y_i$'s to depend on the covariates and allows the random effects to induce dependence between the response values. \citep{a6} further extended the non-standard LRT test to the profile restricted likelihood ratio test (RLRT), focusing on nonparametric mixed models with spline fitting.

Following \citep{a6}, we apply RLRT to test hypothesis (\ref{a_e32}). Under this hypothesis, the RLRT test statistics, $D$, is the deviance of two times the log PRL, $-2l_{PR}(\boldsymbol{\theta})$, i.e. $D=2 l_{PR}(\boldsymbol{\theta})-2 l_{PR}(\boldsymbol{\theta}_0)$. Note that $D$ is the same using either $l_R$ or $l_{PR}$. Assuming that the corresponding regular conditions in \citet{a44} are satisfied for the PRL function model, $D$ converges to
\be\label{a_e33}
D\rightarrow\inf_{\boldsymbol{\theta}\in\tilde{C}_0}\|U-\boldsymbol{\theta}\|^2-\inf_{\boldsymbol{\theta}\in\tilde{C}}\|U-\boldsymbol{\theta}\|^2,
\ee
where $\tilde{C}=\{\tilde{\boldsymbol{\theta}}:\tilde{\boldsymbol{\theta}}=\mathcal{I}(\boldsymbol{\theta}_0)^{T/2}(\boldsymbol{\theta}-\boldsymbol{\theta}_0),\boldsymbol{\theta}\in C_\Omega\}$ is the orthonormal transformation of the cone approximation, $C_\Omega$, of the parameter space $\Omega$ with $\boldsymbol{\theta}_0$ as the vertex, and $\tilde{C}_0=\{\tilde{\boldsymbol{\theta}}:\tilde{\boldsymbol{\theta}}=\mathcal{I}(\boldsymbol{\theta}_0)^{T/2}(\boldsymbol{\theta}-\boldsymbol{\theta}_0),\boldsymbol{\theta}\in C_{\Omega_0}\}$ is the orthonormal transformed cone approximation of the parameter space $\Omega_0$ under the null hypothesis. $U$ is a random vector from $N(0,I)$, and $\mathcal{I}(\boldsymbol{\theta}_0)^{T/2}$ is the right Cholesky square root of p-REML information matrix, i.e. $\mathcal{I}(\boldsymbol{\theta}_0)=\left[\mathcal{I}(\boldsymbol{\theta}_0)\right]^{1/2}\left[\mathcal{I}(\boldsymbol{\theta}_0)\right]^{T/2}$.

Note that under the null hypothesis, $\boldsymbol{\theta}_0=(\lambda^{-1}_x, 0, 0, \rho)^T$, $\rho$ is inestimable. We suggest estimating the parameters with $\rho$ fixed at the average of $\|\mathbf z-\mathbf z'\|^2$ (average on pairwise observations) to not only reduce the parameter space dimensions but also achieve a better fit. Let $\boldsymbol{\theta}=(\lambda^{-1}_x, \lambda^{-1}_z, \lambda^{-1}_{xz})^T=(\theta_1,\theta_2,\theta_3)^T$. Now the cone parameter spaces are reduced to $C_\Omega=[0,\infty)^3$ and $C_{\Omega_0}=[0,\infty)\times\{0\}\times\{0\}$. However, in this problem, all three parameters can be on the boundaries and the orthonormal transformation for the nuisance parameter $\theta_1$ is not invariant, which leads to a transformation for 3 dimensional space. The calculation of (\ref{a_e33}) in a 3 dimensional space becomes considerably more difficult when the information matrix is not diagonal. To simplify the calculation, we consider the special case that $\theta_1\approx0$, which is a reasonable consideration for the Type II diabetes data in a later Section, where the p-REML estimates of $\theta_1$'s are very close to zero for most pathways.

\begin{figure}[bth]
\begin{center}
\includegraphics[width=8.5cm]{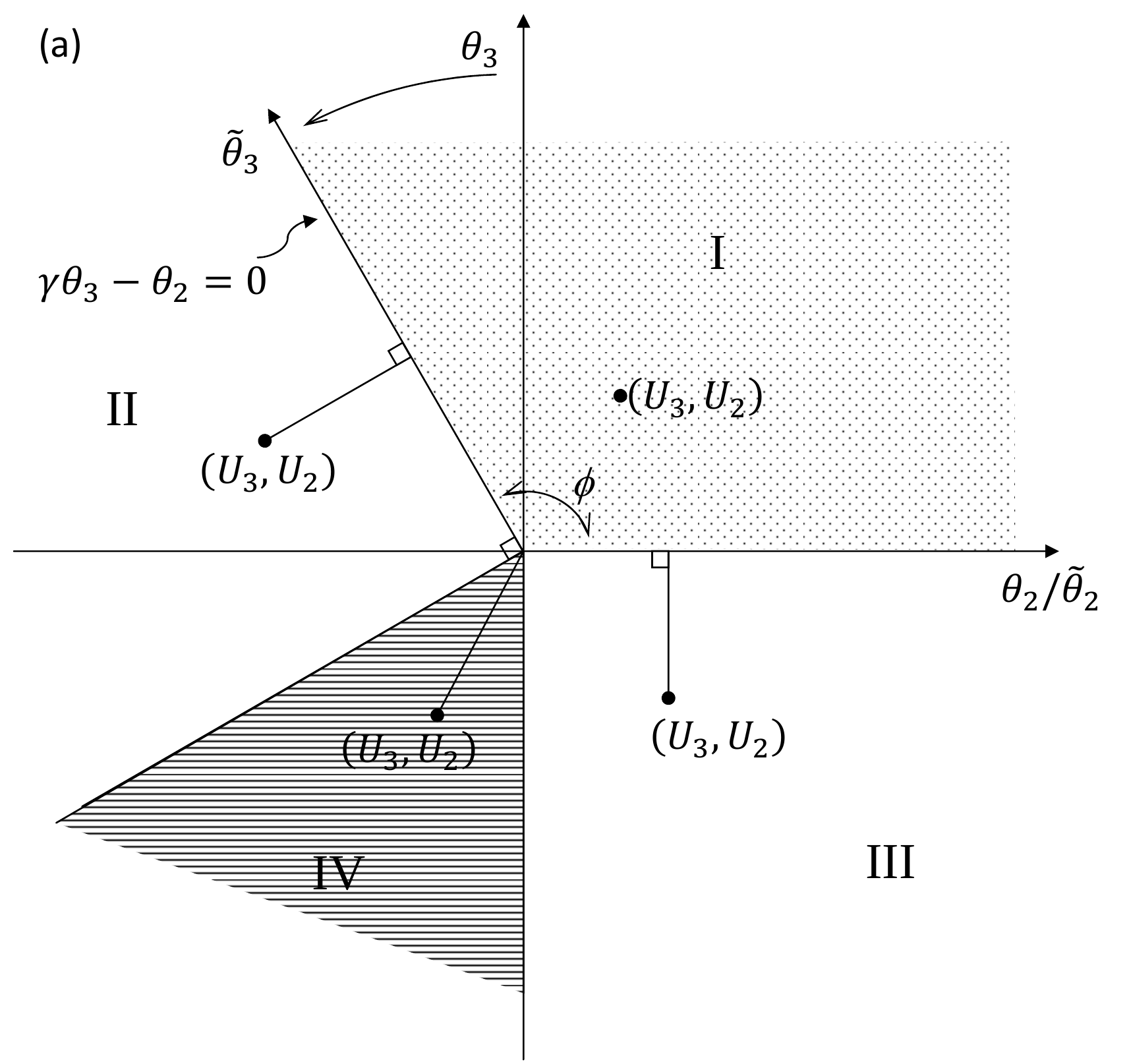}\includegraphics[width=8.5cm]{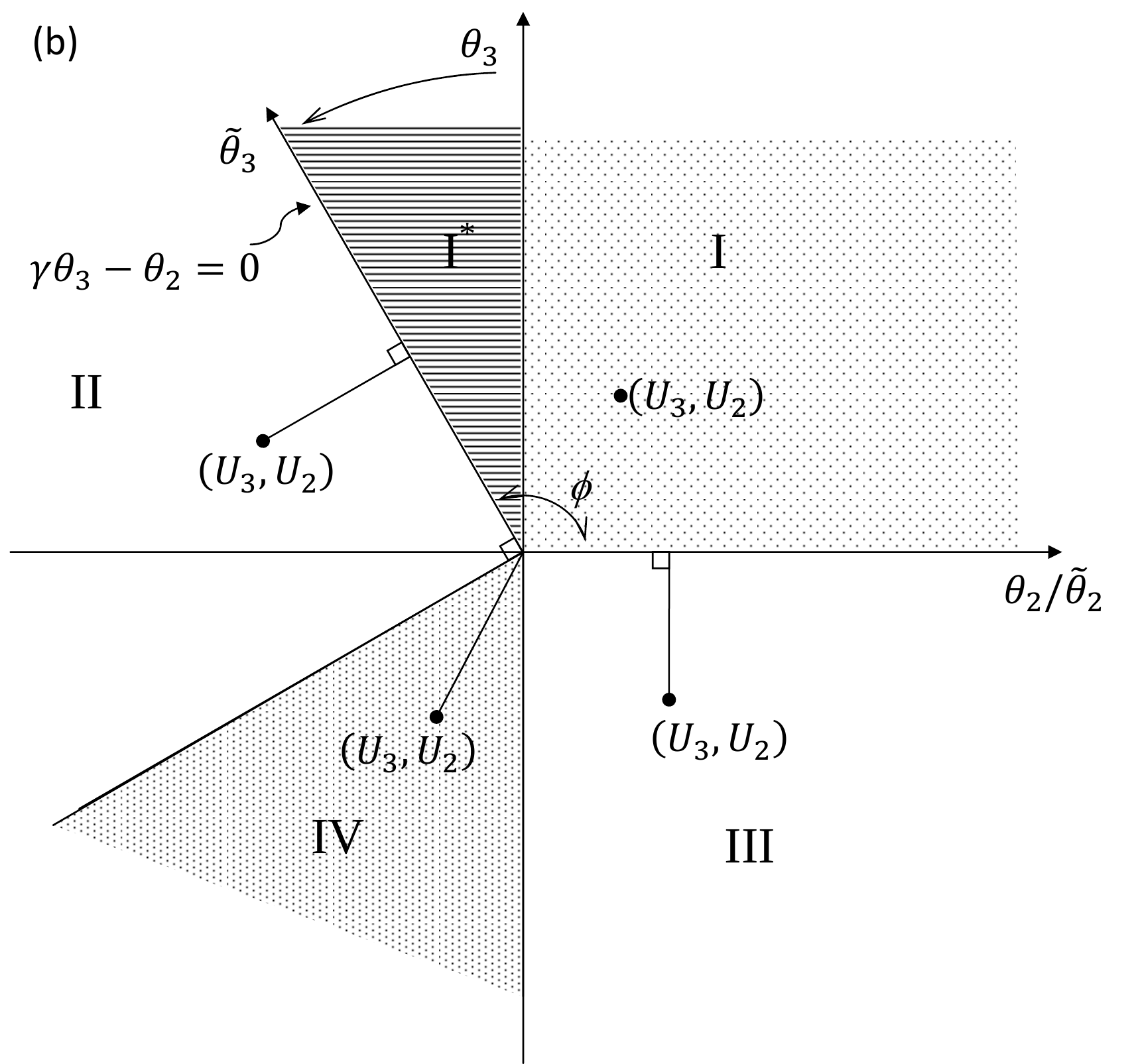}
\caption{Diagram of the parameter space of RLRT for testing two zero variance components (a), and testing the P-E interaction effect (b).}\label{pica}
\end{center}
\end{figure}

Now the parameter space is reduced to 2 dimensions. Under the orthonormal transformation, the cone spaces become to $\tilde{C}=\{\boldsymbol{\theta}: \gamma\theta_3-\theta_2\geq0, \theta_3\geq0\}$, and $\tilde{C}_0=\{\boldsymbol{\theta}: \theta_3=\theta_2=0\}$, where $\gamma=\tilde{\mathcal{I}}_{23}\cdot|\tilde{\mathcal{I}}(\boldsymbol{\theta}_0)|^{-1/2}$ is the slope of the axis $\theta_3$ after transformation as shown in Figure \ref{pica}(a). To account for the fact that $\theta_1$ is estimated, $\tilde{\mathcal{I}}(\boldsymbol{\theta}_0)$ is defined from the $3\times3$ information matrix $\mathcal{I}(\boldsymbol{\theta}_0)$ as
\[\label{a_e34}
\tilde{\mathcal{I}}(\boldsymbol{\theta}_0)=
\begin{bmatrix}  \tilde{\mathcal{I}}_{22} & \tilde{\mathcal{I}}_{23} \\
                 \tilde{\mathcal{I}}_{32} & \tilde{\mathcal{I}}_{33}\end{bmatrix}
=\begin{bmatrix} \mathcal{I}_{22}         & \mathcal{I}_{23} \\
                 \mathcal{I}_{32}         & \mathcal{I}_{33}\end{bmatrix}
-\left[ \begin{array}{c} \mathcal{I}_{21} \\ \mathcal{I}_{31} \end{array} \right]\mathcal{I}^{-1}_{11}\left[\mathcal{I}_{12}, \mathcal{I}_{13}\right].
\]
From the graphic point of view, the representation of the test statistics (\ref{a_e33}) is determined by the minimum distance of the independent normal vector $U=(U_2, U_3)^T$ to $\boldsymbol{\theta}$. Under the alternative hypothesis, the minimum distance, $\inf_{\boldsymbol{\theta}\in\tilde{C}}\|U-\boldsymbol{\theta}\|^2$,  can be understood as the projection of $U$ on the cone space $\tilde C$ when $U$ is outside of the cone. As shown in Figure \ref{pica}(a), the representations of $\inf_{\boldsymbol{\theta}\in\tilde{C}}\|U-\boldsymbol{\theta}\|^2$ are different in the four regions of the plane with coordinates $(\theta_2, \theta_3)$
\be\label{a_e36}
\inf_{\boldsymbol{\theta}\in\tilde{C}}\|U-\boldsymbol{\theta}\|^2=\left\{
\begin{array}{l l l}
  0                                                      & \theta_3\geq0,\;\; \gamma\theta_3-\theta_2\geq0,            &I  \\
  U_2^2+U_3^2-(\gamma U_2+U_3)^2/(1+\gamma^2)            & \theta_3+\gamma\theta_2\geq0,\;\; \gamma\theta_3-\theta_2<0,&II  \\
  U_3^2                                                  & \theta_3<0,\; \theta_2\geq0,                                &III  \\
  U_2^2+U_3^2                                            & \theta_3+\gamma\theta_2<0,\;\; \theta_2<0,                  &IV.  \end{array} \right.
\ee
The area proportions, $(\phi, 1/4, 1/4, 1/2-\phi)$ as in the aforementioned order, of these four regions determine the probabilities that the vector $U$ lies in which region, where $\phi=\cos^{-1}\left(\gamma\cdot(1+\gamma^2)^{-1/2}\right)=\tilde{\mathcal{I}}_{23}\cdot(\tilde{\mathcal{I}}_{22}\tilde{\mathcal{I}}_{33})^{-1/2}$.

Under the null hypothesis, the parameters space is reduced to the origin of the plane, thus according to \citet{a44}
\[\label{a_e35}
\inf_{\boldsymbol{\theta}\in\tilde{C}_0}\|U-\boldsymbol{\theta}\|^2=U^2_2+U^2_3.
\]

Then the asymptotic distribution of $D$ is the difference of the above two representations
\be\label{a_e37}
D\rightarrow\left\{\begin{array}{l l l}
  U_2^2+U_3^2                                     & \textrm{with probability}\;\; \phi,     &I                 \\
  (\gamma U_2+U_3)^2/(1+\gamma^2)                 & \textrm{with probability}\;\; 1/4,      &II                \\
  U_2^2                                           & \textrm{with probability}\;\; 1/4,      &III               \\
  0                                               & \textrm{with probability}\;\; 1/2-\phi, &IV. \end{array} \right.
\ee
Note that because $U_2$ and $U_3$ are independent, thus $(\gamma U_2+U_3)/\sqrt{1+\gamma^2}\sim N(0, 1)$, and the final approximate asymptotic distribution of $D$ is
\be\label{a_e38}
D\sim\phi\chi^2_2+0.5\chi^2_1+(0.5-\phi)\chi^2_0.
\ee
In this paper, we suppose $\lim_{n\rightarrow\infty}|\gamma|<\infty$. If $\lim_{n\rightarrow\infty}|\gamma|\rightarrow\infty$, the representation of $\inf_{\boldsymbol{\theta}\in\tilde{C}}\|U-\boldsymbol{\theta}\|^2$ is in different form \citep{a44} and the asymptotic distribution of $D$ may be different. An additional approximation is that we obtain $\gamma$ with a finite sample size under the null hypothesis, so we assume that $n$ is large enough that the finite $\gamma$ is close to the converged value.
\subsection{Test for the P-E Interaction Effect}\label{a_sec4_2}
The RLRT for two variance components introduced above allows us to test the overall pathway effect. Furthermore, we may be attracted to testing single variance components, such as testing the P-E effect, given that the overall the pathway effect test is significant. The hypothesis of this problem is
\be\label{a_e39}
H_0: \lambda^{-1}_{xz}=0\;\; \textrm{vs.}\;\; H_a: \lambda^{-1}_{xz}>0,
\ee
which is equivalent to testing $H_0: \tau_{xz}=0$ vs. $H_a: \tau_{xz}>0$. The RLRT test statistics $d=2 l_{PR}(\boldsymbol{\theta})-2 l_{PR}(\boldsymbol{\theta}_0)$ for one variance component in semiparametric model with PRL was also suggested by \citet{a6}, and an exact RLRT algorithm was proposed by \citet{a7}. Unfortunately, this exact RLRT method cannot apply to test (\ref{a_e39}) for model (\ref{a_e17}). In their work, there are no random effects in the model under the null hypothesis, thus $d$ can be represented exactly as the form of a mixture of chi-square distribution. On the contrary, our model (\ref{a_e17}) under the null hypothesis of (\ref{a_e39}) contains two random effects $\mathbf r_x$ and $\mathbf{r}_z$, which makes it impossible to represent $d$ exactly.

The second choice is to use the method described in the previous section using an asymptotic distribution. However, we need the same approximations; that is, we fix $\rho$ and assume that the relationship between the response and the environmental variable is almost linear, i.e. $\tau_x\approx0$. Then similarly, the parameters cone space is reduced to 2 dimensions. One interesting parameter $\theta_3=\lambda^{-1}_{xz}$, and one nuisance parameter $\theta_2=\lambda_z^{-1}$, both have the true values on the boundary. Thus, $C_\Omega=[0,\infty)\times[0,\infty)$ and $C_{\Omega_0}=[0,\infty)\times\{0\}$.\\
Under the approximations described above, the asymptotic representation of 2 times the log PRL function under the null hypothesis is
\be\label{a_e40}
\inf_{\boldsymbol{\theta}\in\tilde{C}_0}\|U-\boldsymbol{\theta}\|^2=0\cdot I(U_2>0)+U_2^2I(U_2\leq0)+U_3^2,
\ee
where $I(\cdot)$ is the indicator function. The representation under the alternative hypothesis is the same as in (\ref{a_e36}), but because the cone under the null hypothesis is no longer the origin of the $(\theta_2, \theta_3)$ plane, $\inf_{\boldsymbol{\theta}\in\tilde{C}_0}\|U-\boldsymbol{\theta}\|^2$ has two regions as shown by (\ref{a_e40}). Now we must divide the plane with coordinates $(\theta_2, \theta_3)$ into five regions and set the approximated asymptotic representation of $d$ as (see Figure \ref{pica}(b))
\be\label{a_e41}
d\rightarrow\left\{\begin{array}{l l l}
  U_3^2                                           & \textrm{with probability}\;\; 1/4,        &I   \\
  U_2^2+U_3^2                                     & \textrm{with probability}\;\; \phi-1/4,   &I^*   \\
  (\gamma U_2+U_3)^2/(1+\gamma^2)                 & \textrm{with probability}\;\; 1/4,        &II     \\
  0                                               & \textrm{with probability}\;\; 1/4,        &III    \\
  0                                               & \textrm{with probability}\;\; 1/2-\phi,   &IV.\end{array} \right.
\ee
Thus, we have the asymptotic distribution of $d$ for testing $\theta_3=\lambda_{xz}^{-1}=0$ or $\tau_{xz}=0$
\be\label{a_e42}
d\sim(\phi-0.25)\chi^2_2+0.5\chi^2_1+(0.75-\phi)\chi^2_0,
\ee
where $\phi$ is calculated through $\gamma$ under hypothesis (\ref{a_e39}).

In many cases, the relationship between the response and the environmental variable is not linear, i.e. $\tau_x$ is significant and not equal to 0, then we are in the 3 dimension space to derive the asymptotic distribution of the $d$, which becomes arduous. In this situation, we adopt a score test approach based on the REML function (\ref{a_e22}) which was proposed by \citet{a24} in a mixed model. The asymptotic distribution of the REML score may not converge to a standard normal distribution, \citet{a48}  suggested using the scaled chi-square approximation of the test statistics. More generally, the REML score for covariance component $\tau_\alpha, \alpha\in\{x, z, xz\}$ of (\ref{a_e23}) can also be written as
\[\label{a_e43}
{{\partial l_{R}}\over {\partial\tau_\alpha}}={1\over2}(P\mathbf{y})^T{{\partial \Sigma}\over{\partial\tau_\alpha}}P\mathbf{y}-{1\over2}\Tr\left(P{{\partial \Sigma}\over{\partial\tau_\alpha}}\right),
\]
where we used identity $(\mathbf{y}-X\hat{\boldsymbol{\beta}})^T\Sigma^{-1}=(P\mathbf{y})^T$. $P$ can be expressed as $P=\Gamma(\Gamma^T\Sigma \Gamma)^{-1}\Gamma^T$ \citep{a41}, where $\Gamma^T$ is $(n-q)\times n$ matrix with full row rank $n-q$ ($q=2$ is the rank of $X$). The matrix $\Gamma^T$ satisfies $\Gamma^TX=0$ and $\Gamma^T\mathbf{y}\sim N(0,\Gamma^T\Sigma \Gamma)$. Thus the REML version score test statistics can be written as
\be\label{a_e44}
U_{\tau_\alpha}={1\over2}(P\mathbf{y})^T{{\partial \Sigma}\over{\partial\tau_\alpha}}P\mathbf{y}=\tilde{\mathbf{y}}^TM\tilde{\mathbf{y}},
\ee
where $\tilde{\mathbf{y}}=(\Gamma^T\Sigma \Gamma)^{-{1\over2}}\Gamma^T\mathbf{y}$ with $\tilde{\mathbf{y}}\sim N(0, I_{n-q})$, and $M={1\over2}(\Gamma^T\Sigma \Gamma)^{-{1\over2}}\Gamma^T{{\partial \Sigma}\over{\partial\tau_\alpha}}\Gamma(\Gamma^T\Sigma \Gamma)^{-{1\over2}}$. $U_{\tau_\alpha}$ is the quadratic form of $\mathbf{y}$ with mean $E(U_{\tau_\alpha})={1\over2}\Tr\left(P{{\partial \Sigma}\over{\partial\tau_\alpha}}\right)$ and variance $\var(U_{\tau_\alpha})=\mathcal{I}_{jj}$, where $\mathcal{I}_{jj}$ is the corresponding entry of the information matrix (\ref{a_e24}) for the interesting variance component of $\tau_\alpha\in\{\tau_x, \tau_z, \tau_{xz}\}$.

Let $r$ denote the number of non-zero eigenvalues of $M$, then $M$ can be further decomposed using the spectral decomposition as $M=H\Xi H^T=\sum_{i=1}^r\xi_ih_ih_i^T$, where $H=(h_1,...,h_r)$ is $n\times r$ orthogonal normal matrix, i.e. $h_i^Th_j=\delta_{ij}$, and $\Xi=\langle\xi_i\rangle$ is $r\times r$ diagonal matrix. It follows that
\[\label{a_e45}
U_{\tau_\alpha}=\tilde{\mathbf{y}}^TH\Xi H^T\tilde{\mathbf{y}}=\sum_i^r\xi_i\tilde{\mathbf{y}}^Th_ih_i^T\tilde{\mathbf{y}}\sim\sum_i^r\xi_i\chi^2_1.
\]
Therefore, under $H_0$, the distribution of $U_{\tau_\alpha}$ can be represented as a weighted mixture of chi-square distribution. This is because $\tilde{\mathbf{y}}^Th_ih_i^T\tilde{\mathbf{y}}\sim \chi_1^2$ since $h_ih_i^T$ is an idempotent matrix with rank 1. Because the calculation for $\xi_i$'s is intensive, we follow \citet{a48} in using the Satterthwaite method to approximate the distribution of $U_{\tau_\alpha}$ by a scaled chi-square distribution $\kappa\chi^2_\nu$, where $\kappa=\mathcal{I}_{jj}/2E(U_{\tau_\alpha})$, and $\nu=2E(U_{\tau_\alpha})^2/\mathcal{I}_{jj}$. \citet{a48} also suggested to further account for the fact that $\boldsymbol{\theta}=(\sigma^2, \tau_x, \tau_z, \tau_{xz}, \rho)^T$ is estimated, so that $\kappa$ and $\nu$ are calculated by replacing $\mathcal{I}_{jj}$ with the efficient information $\tilde{\mathcal{I}}{jj}=\mathcal{I}_{jj}-\mathcal{I}_{j\vartheta}\mathcal{I}^{-1}_{\vartheta\vartheta}\mathcal{I}^T_{j\vartheta}$, where $\mathcal{I}_{j\vartheta}$ and $\mathcal{I}_{\vartheta\vartheta}$ are the corresponding vector and matrix if we rearrange the $5\times5$ information matrix $\mathcal{I}(\boldsymbol{\theta})$ as
\[\label{a_e46}
\mathcal{I}(\boldsymbol{\theta})=
\begin{bmatrix}  \mathcal{I}_{jj}           & \mathcal{I}_{j\vartheta} \\
                 \mathcal{I}_{j\vartheta}^T & \mathcal{I}_{\vartheta\vartheta} \end{bmatrix}.
\]
In this paper, we are particularly interested in testing the P-E interaction effect, i.e., $\tau_{xz}$.

\section{Simulation Study}\label{a_sec5}
\subsection{Parameters Estimation}\label{a_sec5_1}
We carried out the simulation study to evaluate the accuracies of the estimators; 200 runs were performed for each of the simulation scenarios. Let $p$ denote the number of genes in the pathway and $n$ denote the number of observations. We considered a setup that mimics the real diabetes pathway data with a total of 50 genes within a pathway. The true model of the $i$th observations is
\[\label{a_e47}
y_i=f_x(x_i)+f_z({\bdz}_i^{T})+f_{xz}(x_i,{\bdz}_i^{T})+\epsilon_i,\,\, \epsilon_i\sim N(0,\sigma^2)
\]
with nonparametric functions
\be\label{a_e48}
\begin{split}
f_x(x_i)&=5.6+0.1x_i+\cos\left({x_i\pi/18}\right),                                                                                \\
f_z({\bdz}_i^{T})&=a\cdot{\bdz}_i^{(30)}\exp\left(-{0.2\bar{|{\bdz}|}_i^{(30)}}\right)/5,                                 \\
f_{xz}(x_i,{\bdz}_i^{T})&=b\cdot e^{x_i/10}\sin\left(\bar{{\bdz}}_i^{(30)}\right)\cos\left(\bar{{\bdz}}_i^{(30)}\right)/8,
\end{split}
\ee
where ${\bdz}_i^{(30)}, \bar{|{\bdz}|}_i^{(30)}$ and $\bar{{\bdz}}_i^{(30)}$ stand for ${\sum_{j=1}^{30}z_{ij}}, \sum_{j=1}^{30}|z_{ij}|/30$ and $\sum_{j=1}^{30}z_{ij}/30$. We sample $x_i$ and $z_{ij} (j=1,...,50)$ from ${\rm Uniform}[18 ,36]$ and $N(0,1)$, respectively. Furthermore, $a$ and $b$ are parameters to control the magnitude of the nonparametric functions respectively. In this Section they are fixed at $a=1.5$ and $b=2$. In the true model (\ref{a_e48}), a total of 30 genes, $z_{i1},...,z_{i30}$, are involved. However in a real situation, we may fit the model with extra genes that are not involved in the true model. Thus we consider the following settings for model (\ref{a_e48}):\\
\indent Setting 1:  $n=100/150$, true $p=30$, fitted $p=30$, $\sigma^2=0.2^2$, \\
\indent Setting 2:  $n=100/150$, true $p=30$, fitted $p=40$, $\sigma^2=0.2^2$, \\
\indent Setting 3:  $n=100/150$, true $p=30$, fitted $p=50$, $\sigma^2=0.2^2$. \\
For each setting, two sample sizes $n=100$ and $150$ were considered.

In Section \ref{a_sec3} we introduced two methods to estimate the variance components using REML and p-REML. We are particularly interested in comparing the performance of these two methods. One of the difficulties of solving equation (\ref{a_e23}) or (\ref{a_e29}) using a scoring method is finding the initial values for $\boldsymbol{\theta}$ or $\boldsymbol{\theta}^*$, since there are no analytic expressions to roughly obtain those initial values. \citet{a3} suggested starting the variance parameters from small positive values within a complex situation. We started the variance components with $(\sigma^2, \tau_x, \tau_z, \tau_{xz})^T=(0.001, 0.001, 0.001, 0.001)^T$, which is equivalent to starting with $(\sigma^2, \lambda_x^{-1}, \lambda_z^{-1}, \lambda_{xz}^{-1})^T=(0.001, 1, 1, 1)$ for p-REML. For scale parameter $\rho$, we can either fix or estimate it. In this simulation study, we choose the initial value $\rho=2$ which is the average of $\|{\bdz}-{\bdz}'\|^2$ on all pairwise observations if it is estimated. We also compare the results with $\rho$ fixed at 2. Note that if $\rho$ is estimated, we consider two possible ways. One way is to perform a two-step procedure where we first fix $\rho$ at 2 and evaluate $(\sigma^2, \tau_x, \tau_z, \tau_{xz})$ until convergence and then use the results with $\rho=2$ as the initial values to evaluate $(\sigma^2, \tau_x, \tau_z, \tau_{xz}, \rho)$ until convergence. The other way is to evaluate $(\sigma^2, \tau_x, \tau_z, \tau_{xz}, \rho)$ together from an initial value $(0.001, 0.001, 0.001, 0.001, 2)^T$. The simulation results show that the former method is more stable, so only these results are shown. Similarly, a two-step procedure was used for p-REML when $\rho$ is estimated.

\begin{figure}[bth]
\begin{center}
\includegraphics[width=15cm]{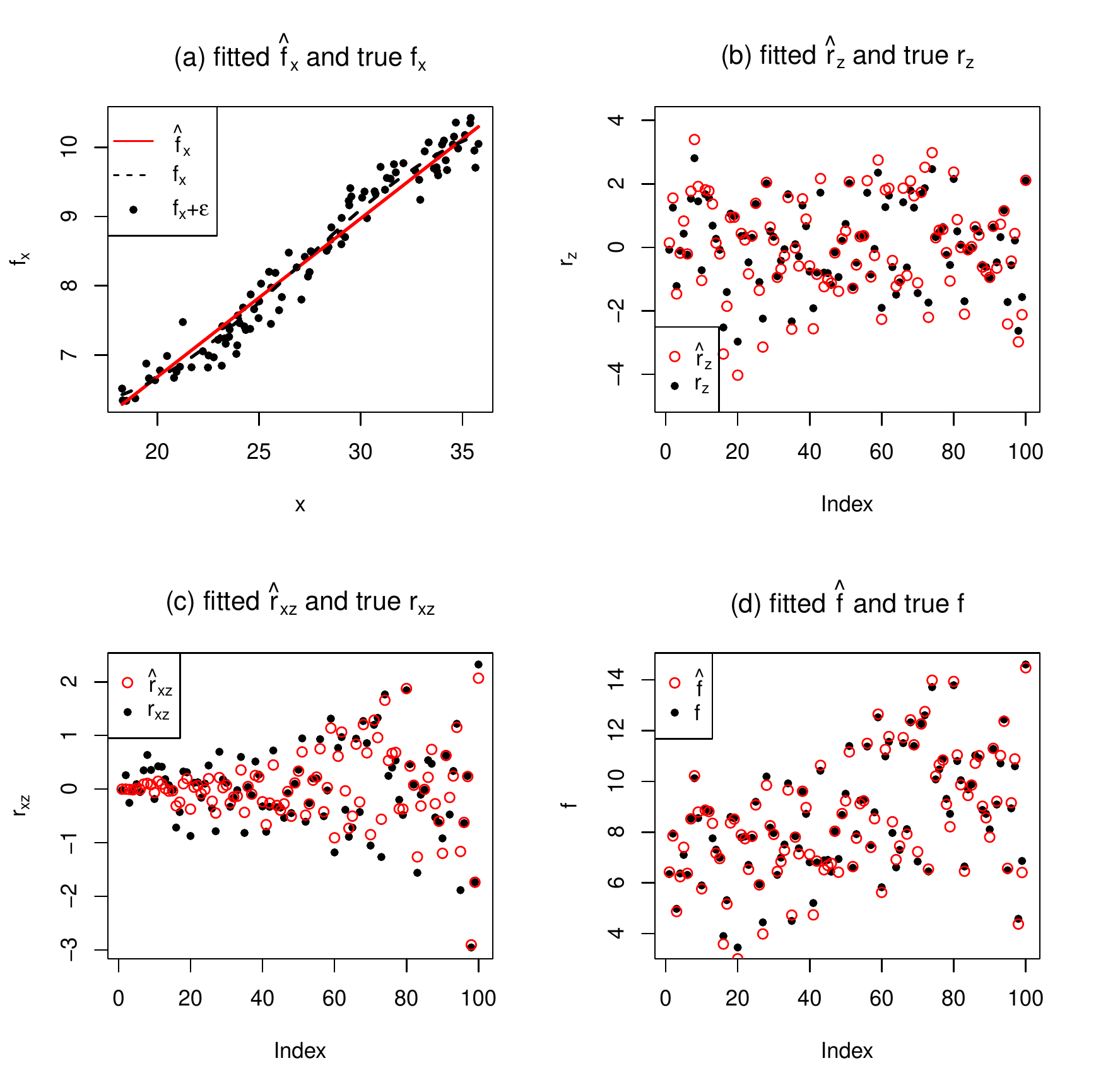}
\caption{Selected example of fitting results of setting 1. Because of the high dimensionality, $\mathbf r_z, \mathbf r_{xz} \hbox{ and }\mathbf f$ are plotted vs. the observation index only.}\label{picb}
\end{center}
\end{figure}

To demonstrate the fitting results, Figure \ref{picb} shows one selected example of setting 1 comparing estimated $\mathbf f, \mathbf f_x, \mathbf r_z\hbox{ and }\mathbf r_{xz}$ with the true ones. The overall response $\mathbf f$ is fitting very well as shown in Figure \ref{picb}(d). As shown in Figure \ref{picb}(b) and (c), there is not much identifiability issue since both the fitted pathway effect $\hat{\mathbf r}_z$ and fitted the interaction effect $\hat{\mathbf r}_{xz}$ capture the patterns of the true ones very well.

To have a overall evaluation of the goodness-of-fit of the nonparametric function $f_x$, $f_z$ and $f_{xz}$, we followed the techniques used by \citet{b17}, who suggested regressing the true nonparametric functions on the fitted ones. By reporting the average intercepts, slopes and $R^2$'s from these regressions, the goodness-of-fit of the fitted nonparametric functions can be assessed empirically. The closer to $0$ and $1$ of the intercepts and slopes are and the closer to $1$ of $R^2$ is, the better the performance of the estimation is.

In Table \ref{table1} we summarized the goodness-of-fit of $f_\alpha, \alpha\in\{x, z, xz\}$ for 200 hundred runs. The scenarios of three settings were used in four procedures: I) REML with $\rho$ estimated, II) REML with $\rho$ fixed at 2, III) p-REML with $\rho$ estimated, and IV) p-REML with $\rho$ fixed at 2. It can be seen that the performance of using procedure I) is not so good; $\rho$ goes to an extremely large value and $f_\alpha$'s deviate from $\hat f_\alpha$'s. This may be because the REML likelihood function dose not have a maximum and the likelihood increases or becomes flat with $\rho$. In such a case, the entries of $K_z$ becomes a matrix of ones. One solution when the REML function becomes flat with $\rho$ is to fix $\rho$ at the turning point of the REML function. In procedure II) we fixed $\rho$ at 2. The average of $\|{\bdz}-{\bdz}'\|^2$ on all pairwise observations is very close to 2 and using this $\rho$ allows us to avoid having extreme values for the entries of $K_z$. The performance of this procedure is improved significantly; all the $R^2$ values are over $90\%$ and close to 1, and the intercepts and slopes of the regressions are close to 0 and 1. However, $\hat\sigma^2$ values are all close to zero. The zero error component happens in REML estimation \citep{a41}, especially with high dimensional parameter spaces.

Table \ref{table1} shows that the performance is much better for the two p-REML procedures. Not only is the fitting of nonparametric functions very good, but the estimate of error variance component $\hat\sigma^2$ is close to the true value. As expected, fitting with extra genes introduces more error, which results in the increase of $\hat{\sigma}^2$. This is because fitting irrelevant genes is equivalent to introducing more noise into the model. However, the results show little difference in fitting $f_\alpha$'s for differently used gene numbers. Increasing the number of observations is expected to improve the fitting performance. Although overall there is no much difference between $n=100$ and 150, there is slight improvement in fitting the P-E interaction effect. This can be seen from the fact that $R^2$ increases and the slope of regressing $f_{xz}$ on $\hat{f}_{xz}$ is closer to 1 for $n=150$.

The overall goodness-of-fit using p-RMEL is very good, except there are small biases: the regression slope of $f_z$ on $\hat{f}_z$ is slightly smaller than 1 and the the regression slope of $f_{xz}$ on $\hat{f}_{xz}$ is slightly larger than one. This means that $f_z$ is overestimated and $f_{xz}$ is slightly underestimated. However, for each $f_z$ and $f_{xz}$, the fitted results can explain most of the variations as all the $R^2$ values are very close to 1. We also realized that the fitting of $f_z+f_{xz}$ is much better than individual ones (the regression parameters of $f_z+f_{xz}$ on $\widehat{{f_z+f_{xz}}}$ are not shown), which is easy to be understood if we can treat $\mathbf{r}=\mathbf{r}_z+\mathbf{r}_{xz}$ as one random effect with covariance $\tau_zK_z+\tau_{xz}K_{xz}$. This indicates that there is no bias in fitting $f_z+f_{xz}$, but the weight between $f_z$ and $f_{xz}$ might be biased. The reason for this can be understood from the interaction kernel expression ($\ref{a_e9}$). It can be seen that if the entries of matrix $xx'+k_x(x,x')$ are close to each other, then $\tau_zK_z+\tau_{xz}K_{xz}$ is nothing more than a scalar times $K_z$, and we will have overestimation of $f_z$. However, this bias is not too significant, because the good fit of $f_z+f_{xz}$ and the high $R^2$ values of fitting $f_{xz}$ indicate that it has little influence on testing either the overall pathway or the P-E interaction effect.
\subsection{Test Study}\label{a_sec5_2}
To obtain better convergence, for the rest of this paper we adopt the Marquardt procedure as a scoring method. With the Marquardt method we have flexible iteration steps, this is
\[\label{a_e49}
\boldsymbol{\theta}^{(k+1)}=\boldsymbol{\theta}^{(k)}+\left[\mathcal{I}(\boldsymbol{\theta}^{(k)})+\delta^{(k)}I\right]^{-1}\left.{{\partial l_{R}}\over{\partial \boldsymbol{\theta}}}\right|_{\boldsymbol{\theta}^{(k)}},
\]
where $l_R,\, \mathcal{I}$, and $\boldsymbol{\theta}$ are replaced by the counterparts of the p-REML procedure when it is required. The scalar $\delta^{(k)}$ partially determines the step size and $I$ is the identity matrix. If $\delta^{(k)}$ is small, the procedure approximates a scoring method. If $\delta^{(k)}$ is large, a small step is taken in approximately the direction of the scoring method. We modify $\delta^{(k)}$ accordingly to achieve increasing likelihood. In this paper, we start from $\delta^{(k)}=(1.0\times10^{-5})\times{{\Tr\left(\mathcal{I}(\boldsymbol{\theta})\right)}\over{\textrm{number of }\theta_i'\textrm{s}}}$ to make the initial step size as large as possible.

We first studied the performance of RLRT of testing two zero variance components under hypothesis (\ref{a_e32}). In this simulation study we are particularly interested in two issues: how RLRT performs at different fixed $\rho$ values since we prefer to estimate the parameters with $\rho$ fixed and how the performance degrades with irrelevant genes included in the model. The true model used and the data generating method are the same as described for (\ref{a_e48}) in Section \ref{a_sec5_1}. For both issues, we first set $a=0$ and vary $b$, and then set $b=0$ and vary $a$. It turns out the test is very powerful when both $a$ and $b$ are not equal to zero, so no simulation on this situation is shown here. For all cases, the total number of simulation runs is 1000 times. In addition, the function $f_x(\cdot)$ in (\ref{a_e48}) has a trivial nonlinear component, so we can apply RLRT in this simulation.

For the first issue, we consider the case where the sample size is $n=100$, and both the true and used gene numbers are $p=30$. Table \ref{table2} presents the Type II errors and powers of testing hypothesis (\ref{a_e32}) for 2 sets of $\{a, b\}$ values at 4 different $\rho$ values (one is estimated). In general, the power curve of RLRT does not depend on $\rho$ very much. \citet{b17} revealed the same phenomena for the score test of a single variance component within a model with only one random effect. This is because moderate differences of $\rho$ do not change the structure of the covariance matrix very much, except for extreme values such as $\rho\rightarrow0$ or $\rho\rightarrow\infty$, with which the covariance matrix turns to an identity matrix or a matrix of ones. Note that the empirical Type II errors of all situations are around 0.03, smaller than the nominal one. The reason could be the approximation of (\ref{a_e38}) due to the assumption, $\theta_1=\lambda_x^{-1}\approx0$.

To test two zero variance components with extra genes, we consider simulations with the sample sizes $n=60$ and $n=35$. The latter mimics the Type II diabetes data where the total subjects under study are $n=35$. Fitting with the equal true and used gene numbers is compared to fitting with an extra 20 irrelevant genes. The results in Table \ref{table3} show that, when fitting with extra genes, the power decreases as expected but not dramatically, which means that  the model we proposed can be applied to pathway data for which only some of the genes are related to the responses. In addition, comparing Table \ref{table2} and \ref{table3} shows that the power does decrease with the sample size $n$.

The simulation study for testing P-E interaction using RLRT and the score test is carried out using a new setup for the data  generation. We continue using the same nonparametric expression (\ref{a_e48}) except with true gene number $p=5$; that is, simply replacing $f_z(\cdot)$ and $f_{xz}(\cdot)$ as $f_z({\bdz}^T_i)=a\cdot{\bdz}_i^{(5)}\exp\left(-{0.2\bar{|{\bdz}|}_i^{(5)}}\right)/5$ and $f_{xz}(x_i,{\bdz}^T_i)=b\cdot e^{x_i/10}\sin\left(\bar{{\bdz}}_i^{(5)}\right)\cos\left(\bar{{\bdz}}_i^{(5)}\right)/8$, where ${\bdz}_i^{(5)}=\sum_{j=1}^{5}z_{ij}$, $\bar{|{\bdz}|}_i^{(5)}=\sum_{j=1}^{5}|z_{ij}|/5$ and $\bar{{\bdz}}_i^{(5)}=\sum_{j=1}^{5}z_{ij}/5$.  $x_i, z_{ij}$ and $\epsilon_i$ are generated the same way as before. Note the function form changes when the gene number is different in (\ref{a_e48}). We use this setup to compare two test procedures for testing (\ref{a_e39}). For the score test, we first estimate the parameters using p-REML and then calculate the statistics using expressions (\ref{a_e24}) and (\ref{a_e44}). The results are listed in Table \ref{table4}. Again, we see that the test's power does not depend on $\rho$. The results indicate that the RLRT are slightly lower in power and that the type I errors of the two test methods are all closer to the nominal $5\%$ from different directions. These results indicate we can apply both test methods under suitable conditions.
\section{Application to Type II Diabetes Data}\label{a_sec6}
We applied our mixed model (\ref{a_e17}) to a set of diabetes data from \citet{b21}. They utilized the HGC-133a Affymetrix genechip with 22,283 genes to study 17 normal glucose tolerance individuals vs. 18 Type II diabetes mellitus patients. The 22,283 genes make up a total of 251 pathways. The goal of this study is to identify pathways with the highest significant overall pathway effect when an environmental variable, body mass index, is present in the model, and from them identify pathways with significant P-E interaction effect. Therefore, there are a total of 251 sets of data, each having $n=35$ observations. Corresponding to each individual pathway, the data set contains $(\mathbf{y}, X, Z)$, where $\mathbf{y}$ is the outcomes of glucose level, $X$ has the same meaning as before with the first column of 1's and the second column as the body mass index data of 35 subjects, and $Z (n\times p)$ is the gene expression levels of each pathway,  which contains the number of genes ranging from $p=3$ to $p=543$.

The fitting results of the top 20 pathways are listed in Table \ref{table5} ranked ascendingly in the $p$-value of testing the overall pathway effect using RLRT $D$. It has almost an identical order of the magnitude as the $D$. It can be seen that 19 out of the 251 pathways are significant. For each pathway, the variance components are estimated using p-REML methods and the standard error of those parameters including $\hat{\sigma}^2$ are calculated using information matrix (\ref{a_e24}) with the p-REML estimates plugged in. Again, the initial values for the variance parameters are $(\sigma^2,\lambda_x^{-1},\lambda_z^{-1},\lambda_{xz}^{-1})^T=(0.001,1,1,1)^T$ and $\rho$ is fixed at the average of $\|\mathbf{z}-\mathbf{z}'\|^2$ of different pairwise observations, which ranges from 0.1 to 1.8 for different pathways.

\begin{figure}[bth]
\begin{center}
\includegraphics[width=15cm,totalheight=15cm]{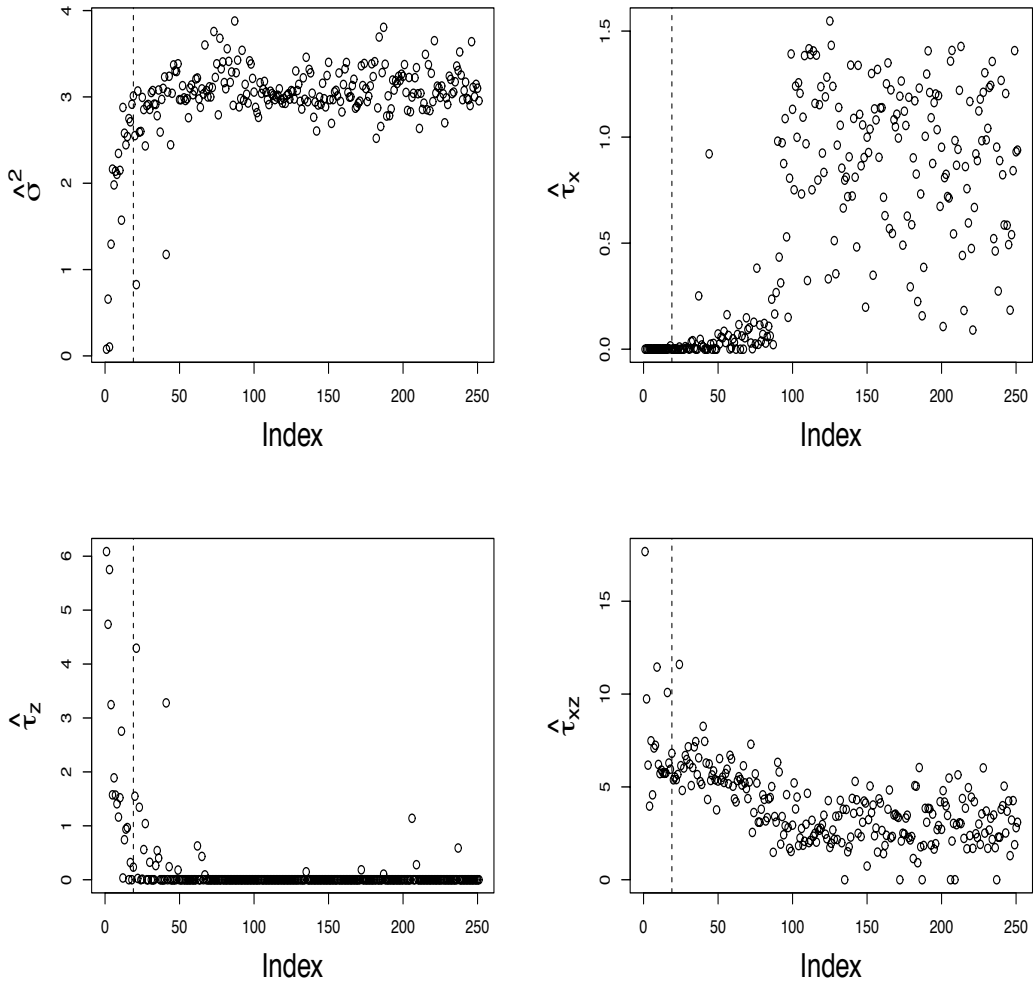}
\caption{The estimated variance components of $\hat\sigma^2, \hat\tau_x, \hat\tau_z, \hat\tau_{xz}$ for 251 pathways ordered by $p$-values of testing the overall pathway effect. The dash lines separate the significant and insignificant pathways at 5\% level.}\label{pic1}
\end{center}
\end{figure}

To show an overall view of the fitting results for 251 pathways, Figure \ref{pic1} plots the four estimated variance components in the same order of the $p$-value of RLRT $D$. The straight dashed line divides the significant and insignificant pathways of RLRT. The error components, $\hat\sigma^2$'s, are around the constant 3.0 except for those top significant pathways. This is consistent with the test results indicating that for those pathways with genes relevant to the responses, the error is reduced since part of the variation of the responses is explained by pathway main effect or P-E interaction effect. The variations of $\hat{\tau}_x$ and $\hat{\tau}_z$ seems to compensate for each other. For the top 50 pathways, $\hat{\tau}_x$'s are close to zero and $\hat{\tau}_z$ values are large. On the other side, for those pathways which are ranked as lower than 50, $\hat{\tau}_z$ values are very small and $\hat{\tau}_x$ values increase. This indicates that for those pathways not relevant enough to the response, part of the variation of response is explained by the nonlinear relationship of the responses and the environmental variable.  The variation of $\hat\tau_{xz}$ seems less dramatic than other random effects. It does not decrease to zero for those non significant pathways, and stabilizes after the top 100 pathways. {However, using the test of RLRT $d$, we show that the lower ranked pathways, ranked as [50, ..., 251], are not significant in the interaction effect.} These results suggest that the body mass index is important in explaining the relationship between the glucose level and the genetic pathway since many pathways that are significant in the overall pathway effect are either significant in the interaction effect or not.

Because the distribution for $D$ is asymptotic, the $p$-value calculated based on 35 observations may not be as accuracte as expected. Hence, we carried out a permutation test process to obtain the exact distribution of $D$ as follows:
\begin{itemize}
\item \emph{Step 1}: We fit the observed data with the full model (\ref{a_e17}) and reduced model under hypothesis (\ref{a_e32}) using the p-REML approach. In both models, we set $\tau_x=0$ since we assume that $\tau_x$ is insignificant when deriving (\ref{a_e38}). Then we obtained test statistics $D$, and calculated the residual $\hat{\boldsymbol\epsilon}_{0}=\hat{\mathbf r}_z+\hat{\mathbf{r}}_{xz}+\hat{\boldsymbol\epsilon}$ using the fitted results of the full model from $\mathbf y=X\boldsymbol\beta+\mathbf r_z+\mathbf r_{xz}+\boldsymbol\epsilon$.
\item \emph{Step 2}: We permuted the residual $\hat{\boldsymbol\epsilon}_0$ to get new $\hat{\boldsymbol\epsilon}_0^*$ and simulate outcomes as $\mathbf y^*=X\hat{\boldsymbol\beta}+\hat{\boldsymbol\epsilon}_0^*$.
\item \emph{Step 3}: Based on $\mathbf y^*$, $X$ and $Z$, we fit the full model (\ref{a_e17}) and reduced model under hypothesis (\ref{a_e31}) again using the p-REML approach and then calculated the test statistics $D^*$.
\item \emph{Step 4}: We repeated Steps 2-3 for a large number of times (e.g. 10,000 times).
\item \emph{Step 5}: We obtained the empirical $p$-value of the RLRT by formula $p$-value = (number of $D^*$'s greater than $D$) $\div$ (total number of $D^*$'s).
\end{itemize}
The $p$-value of the permutation test of $D$ as well as the RLRT $D$ are listed in Table \ref{table6} in the same order of Table \ref{table5} for the top 20 pathways. Note that for RLRT if the sample size is too small such that the information matrix (\ref{a_e30}) is non positive definite, $\phi$ in (\ref{a_e38}) cannot be calculated, so we are not able to get the asymptotic distribution of $D$. However the information matrices of the 251 pathways under hypothesis (\ref{a_e32}) are all positive definite (not true under hypothesis (\ref{a_e39})), so we are able to test the overall pathway effect for all using RLRT $D$.  The results of both tests are similar to each other with respect to the general rank of the significance, specifically both tests have the same top 3 pathways, which are pathways 73, 274, and 230. 
In addition, most of the $p$-values of the permutation tests are slightly larger than those of RLRT, as expected, since the permutation test is usually more conservative. Table \ref{table6} also labels those significant pathways ranked in the top 50 list according to the global score test \citep{a10} and the forest tree method \citet{a35,a36}, which do not take into account the environmental variable in their models. 
Our approach identified pathways that have either significant main pathway effect, the interaction effect, or both, while other methods determined many as having a significant main pathway effect only. Through following one zero variance component test, we also discovered that some pathways have a significant P-E interaction effect although they may not have a significant main pathway effect.

\begin{figure}[bth]
\begin{center}
\includegraphics[width=15cm,totalheight=15cm]{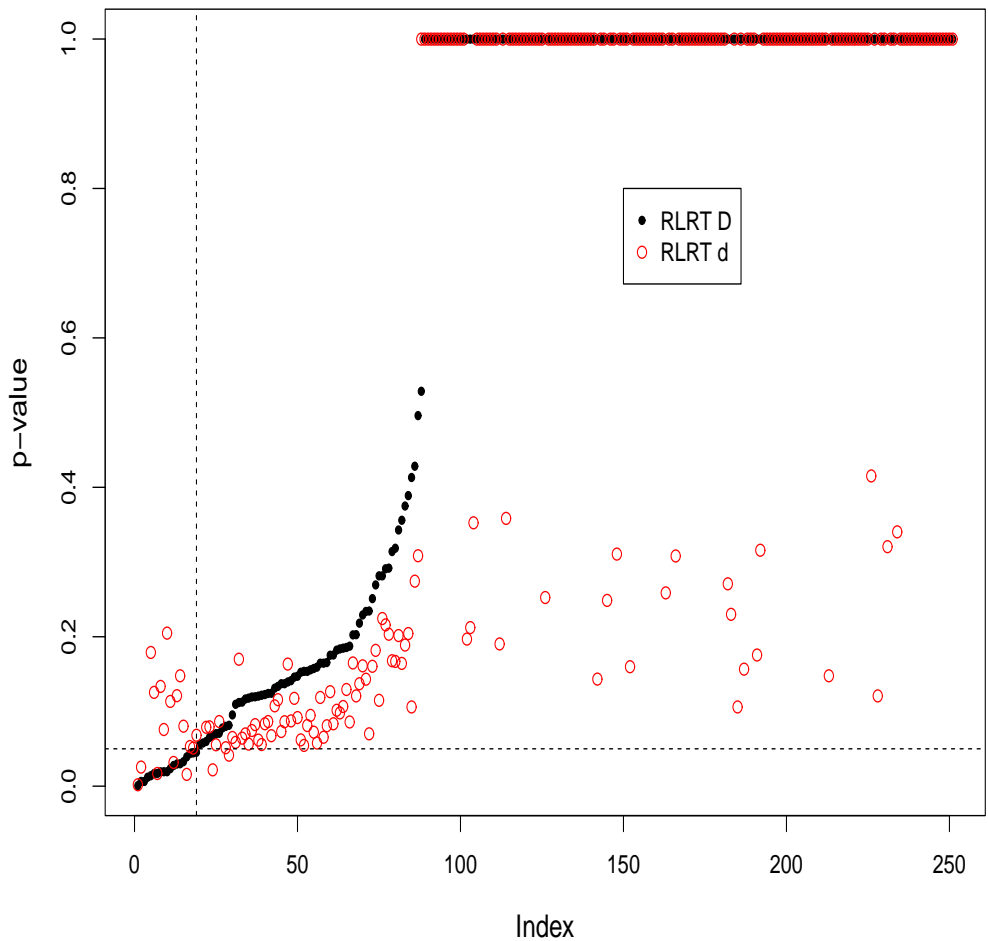}
\caption{The $p$-values of testing overall pathway effect (RLRT $D$) and P-E interaction effect (RLRT $d$) for 251 pathways. The vertical dash line divides the significant and insignificant pathways of overall pathway effect test, and the horizontal dash line indicates 5\% significant level. Some $p$-values of RLRT $d$ are missing because the information matrix is not positive definite.}\label{pic2}
\end{center}
\end{figure}

Furthermore, the $p$-values of RLRT $d$ are also listed in Table \ref{table6}. There are pathways for which we are unable to calculate $d$ because the information matrix is not positive definite. In Figure \ref{pic2} the $p$-values of RLRT $D$ and RLRT $d$ of all pathways are plotted for comparison. Among the top 50 that are significant in overall pathway effect, only part of them are significant in the interaction effect, but for the remaining 151 pathways, none are significant in either interaction effect or overall pathway effect. Similar to RLRT $D$, a permutation test process for the exact distribution of RLRT $d$ is introduced here:
\begin{itemize}
\item \emph{Step 1}: We fit the observed data with the full model (\ref{a_e17}) and reduced model under hypothesis (\ref{a_e39}) using the p-REML approach. Again in both models we assume that $\tau_x$ is negligible. Then we obtained $d$, and calculated the residual $\hat{\boldsymbol\epsilon}_{0}=\hat{\mathbf{r}}_{xz}+\hat{\boldsymbol\epsilon}$ using the fitted results of the full model from $\mathbf y=X\boldsymbol\beta+\mathbf r_z+\mathbf r_{xz}+\boldsymbol\epsilon$.
\item \emph{Step 2}: We permuted the residual $\hat{\boldsymbol\epsilon}_0$ to get new $\hat{\boldsymbol\epsilon}_0^*$ and simulated outcomes as $\mathbf y^*=X\hat{\boldsymbol\beta}+\hat{\mathbf r}_z+\hat{\boldsymbol\epsilon}_0^*$.
\item \emph{Step 3}: Based on $\mathbf y^*$, $X$ and $Z$, we fit the full model and reduced model under hypothesis (\ref{a_e39}) again using the p-REML approach and then calculated the test statistics $d^*$.
\item \emph{Step 4}: We repeated Steps 2-3 a large number of times (e.g. 10,000 times).
\item \emph{Step 5}: We obtained the empirical $p$-value of the RLRT by formula $p$-value = (number of $d^*$'s greater than $d$) $\div$ (total number of $d^*$'s).
\end{itemize}
The permutation test results of RLRT $d$ are close to those of RLRT $d$ in the 20 pathways, but it is difficult to tell which one is more conservative.

We also calculated the $p$-values of testing $H_0$ (\ref{a_e39}) using the score test approach for the top 20 pathways. Compared with the RLRT $d$ and RLRT $d$ permutation tests, the $p$-values of the score test is similar in sense of determining the significant pathways at the 5\% level. Among these top 20 pathways with significant overall pathway effect, the pathways with insignificant interaction effect are $\{229, 152, 16, 236, 144, 151, 103, 271, 101, 158\}$ according to the score test, and $\{229, 152, 16, 236, 144, 151, 14, 103, 271, 150, 158\}$ according to the RLRT $d$ permutation test. Note that the difference of the two sets, $\{14, 101, 150\}$,  all have marginal $p$-values for the two tests at the $5\%$ level. If they are removed from the two sets, both tests have identical pathways which have insignificant P-E environment interaction effects.

Based on the three tests procedures, we identified the pathways with a significant P-E environment interaction effect for all tests among the top 20 pathways. They are $\{ 73, 274, 230, 173, 228, 172\}$ pathways at the 5\% level. These pathways are known to be  related to Type II diabetes. 
Pathway 73 is a {\it Cysteine metabolism} pathway. It is known that taurine (a semi-essential sulphur amino acid) derived from cysteine metabolism can prevent diabetes mellitus and/or insulin resistance \citep{a9}. Pathway 274 is  involved in the {\it Urea cycle and metabolism of amino groups}, which has also been reported to be related to Type II diabetes \citep{a8}. Pathway 230 is {\it OXPHOS\_HG-U133A\_probes} pathway. It has been reported that genes involved in oxidative phosphorylation are coordinately upregulated with fasting hyperglycaemia in the livers of patients with Type II diabetes \citep{a31}. The transcription levels of a class of genes involved in oxidative phosphorylation mechanisms are consistently lower in diabetics than in controls \citep{b21,a31}. Pathway 173 is $MAP00531\_Glycosaminoglycan\_degradation$ pathway. It is known that Type II diabetes mellitus also induces an increased urinary excretion of total glycosaminoglycans \citep{a21}. Pathway 228 is involved in $Oxidative\, phosphorylation$. It is known to be related to diabetes \citep{a31,b21,a34}. This pathway is a process of cellular respiration in humans (or in general eukaryotes) and contains coregulated genes across different tissues and is related to insulin/glucose disposal. It is associated with ATP synthesis, a pathway involved in energy transfer. Pathway 172 is $MAP00530\_Aminosugars\_metabolism$ pathway. Aminosugars (= glucosamine) have no effect on fasting blood glucose levels, glucose metabolism, or insulin sensitivity at any oral dose level in healthy subjects, individuals with diabetes, or those with impaired glucose tolerance \citep{a42}.

\section{Discussion}\label{a_sec7}
The development of a pathway-based mixed model to relate the response with genetic pathways is motivated by the fact that genes always interact with the environmental variables. Modeling the P-E interaction effect can help in further understanding the biological mechanisms underlying diseases and facilitate the discovery of potential biomarkers. However, no existing approaches are able to jointly analyze pathways with the environmental variables when P-E interaction exists.

In this paper, we have addressed a mixed effects model connecting with kernel machine methods and smoothing spline, so that we can analyze the genetic pathway data with a continuous clinical outcome when the P-E interaction effect is present in the model. We demonstrated the application of our method to a pathway data of Type II diabetes. Our approach allows us to evaluate the pathway effect and its interaction with the environmental variables by estimating the corresponding variance components and testing the significance of those parameters.
Because of the high dimensional parameters space, there are usually some difficulties in solving the REML equations, such as non-positive error estimated. We reduced the parameter space dimension in solving REML equations by introducing the p-REML approach to estimate the variance components so that the error component is always in the parameter space. The p-REML approach not only allows us to solve the REML equations efficiently, but also provides an efficient choice in testing one or two zero variance components besides the global score test, i.e. the profile restricted likelihood ratio test for testing the overall pathway effect or P-E interaction.

Modeling the linear mixed model with a kernel machine has other advantages. It allows us to choose appropriate kernels to construct the variance matrix of the random effect as well as the interaction random effect in accordance with the data structure. In this paper, we focused on the Gaussian kernel, but when the sample size is large so that the computation becomes expensive, some less computational intensive alternatives to Gaussian kernel are available, such as rational quadratic kernel: $k({\bdz}^T,{\bdz}'^T)=1-\|{\bdz}-{\bdz}'\|^2/(\|{\bdz}-{\bdz}'\|^2+c)$. Other kernels, such as a polynomial kernel, an exponential kernel, an inverse multiquadric kernel, etc., have also been examined and can replace the Gaussian kernel in appropriate situations. Note that these kernels are similar to the Gaussian kernel in terms of reducing the dimension of the covariates through measuring the similarity of ${\bdz}$ and ${\bdz}'$. To some extent, this may be a disadvantage of the kernel method since there may be some information lost beyond the similarity of the two attributes.

Possible extensions of our method include applying the interaction kernel machine to generalized linear models. Logistic kernel machine regression with a Gaussian kernel has been developed by \citet{a27}, but no interaction between the genetic pathway effect and environmental variable has been considered. By adding the interaction kernel machine to a generalized linear model, our method can be applied in more general genomewide association studies, especially in the case-control studies of G/P-E interaction. The second potential extension of our method is to consider a higher dimension of environmental variables ${\bdx}_i^T$, such as bivariate ${\bdx}_i^T=(x_{i1}, x_{i2})$, longitude and latitude data, and the nonparametric function $f_x({\bdx}^T_i)$ can be fitted using thin plate splines \citep{a15}. With the kernel of the thin plate splines, we can construct the interaction function space kernel similarly. This extension may have wider applications such as in spatial data where the interaction between location and other high dimensional covariates are particularly interesting.

We note that we evaluate the interaction between each pathway and environmental variable. It is known that pathways are not independent of each other because of shared genes and interactions among pathways as well as their interaction with environmental variables, making it difficult to adjust the $p$-value due to the complex dependency structure. Because existing multiple comparison methods based on false discovery rates \citep{a2,a43} were developed only for single gene based analysis that did not take into account the interaction between genes and environmental variables, they are not applicable in such a complicated situation as our problem. Developing a multiple comparison method will be an interesting and challenging problem because of the complex dependence structure among pathways and environmental variables.

\section*{Acknowledgements}
This study was supported in part by the National Science Foundation
grant number 0964680.
\pagebreak

\pagebreak\clearpage\newpage
\thispagestyle{empty}
\begin{table}[h]
\centering
\caption{Assessments of estimating $f_x, f_z$ and $f_{xz}$ simulated by (\ref{a_e48}) using REML and p-REML procedures with $\rho$ estimated from initial value 2 or fixed at 2. Total runs number 200 for each scenario, and the average values are reported.}
\scriptsize
\begin{tabular}{cccccrccrccrcc}
\hline\hline
   &                 &fitted $p$&            &$\hat{\rho}$&\multicolumn{3}{c}{$f_x\sim\hat{f}_x$}&\multicolumn{3}{c}{$f_z\sim\hat{f}_z$}&\multicolumn{3}{c}{$f_{xz}\sim\hat{f}_{xz}$}\\\cline{6-14}
   &          $n$    &(true $p$)& $\hat{\sigma}^2$ &(initial $\rho$)&Int&Slope&$R^2$&Int         &Slope           &$R^2$  &Int&Slope&$R^2$ \\
\hline
         &\multirow{3}{*}{100} & 30(30)& 0.34& 2130(2)&-0.38&1.00&0.97&-0.01&10.51&0.90&-0.14&5.19&0.46\\
         &                     & 40(30)& 0.29& 1824(2)&-0.55&1.06&0.96& 0.01&11.65&0.89&-0.11&4.17&0.50\\
REML     &                     & 50(30)& 0.32& 1929(2)&-1.53&1.26&0.96&-0.02&16.07&0.87&-0.13&5.28&0.48\\\cline{2-14}
$\rho$   &\multirow{3}{*}{150} & 30(30)& 0.26& 1604(2)&-1.15&1.17&0.98& 0.09& 5.87&0.93&-0.17&3.70&0.54\\
estimated&                     & 40(30)& 0.29& 1814(2)&-0.68&1.18&0.97&-0.09& 8.65&0.91&-0.15&3.95&0.48\\
         &                     & 50(30)& 0.35& 2054(2)&-1.24&1.18&0.97& 0.06&12.32&0.90&-0.15&4.79&0.45\\
\hline
         &\multirow{3}{*}{100} & 30(30)& 6.9e-10& 2   & 0.10&0.99&0.99& 0.01&0.85&0.99& 0.01&1.44&0.90\\
         &                     & 40(30)& 8.6e-10& 2   & 0.13&0.98&0.98& 0.02&0.86&0.98& 0.00&1.40&0.90\\
REML     &                     & 50(30)& 8.5e-10& 2   & 0.16&0.98&0.96& 0.01&0.86&0.98& 0.01&1.41&0.88\\\cline{2-14}
$\rho$   &\multirow{3}{*}{150} & 30(30)& 8.5e-10& 2   & 0.05&0.99&0.99& 0.01&0.84&0.99& 0.01&1.41&0.93\\
fixed    &                     & 40(30)& 8.7e-10& 2   &-0.00&1.00&0.99& 0.00&0.84&0.99&-0.01&1.40&0.92\\
         &                     & 50(30)& 7.1e-10& 2   & 0.10&0.99&0.99& 0.02&0.85&0.99& 0.00&1.38&0.91\\
\hline
         &\multirow{3}{*}{100} & 30(30)& 0.04& 3.96(2)&-0.24&1.04&1.00& 0.01&0.85&0.99&-0.04&1.38&0.90\\
         &                     & 40(30)& 0.07& 3.36(2)&-0.19&1.03&1.00&-0.01&0.87&0.99&-0.05&1.46&0.89\\
p-REML   &                     & 50(30)& 0.09& 4.72(2)&-0.31&1.04&1.00& 0.06&0.90&0.98&-0.04&1.44&0.88\\\cline{2-14}
$\rho$   &\multirow{3}{*}{150} & 30(30)& 0.02& 3.00(2)&-0.28&1.04&1.00& 0.01&0.85&0.99&-0.05&1.29&0.92\\
estimated&                     & 40(30)& 0.02& 3.63(2)&-0.29&1.04&1.00& 0.01&0.86&0.99&-0.04&1.29&0.91\\
         &                     & 50(30)& 0.04& 3.19(2)&-0.13&1.02&1.00& 0.01&0.85&0.99&-0.02&1.37&0.91\\
\hline
         &\multirow{3}{*}{100} & 30(30)& 0.04& 2     &-0.08&1.01&1.00& 0.02&0.85&0.99&-0.01&1.64&0.91\\
         &                     & 40(30)& 0.11& 2     &-0.17&1.03&0.99&-0.00&0.88&0.98&-0.03&1.52&0.91\\
p-REML   &                     & 50(30)& 0.11& 2     &-0.12&1.02&0.99&-0.00&0.90&0.98&-0.01&1.38&0.91\\\cline{2-14}
$\rho$   &\multirow{3}{*}{150} & 30(30)& 0.02& 2     &-0.08&1.01&1.00& 0.02&0.86&0.99&-0.01&1.34&0.93\\
fixed    &                     & 40(30)& 0.03& 2     &-0.11&1.02&1.00&-0.01&0.85&0.99&-0.05&1.37&0.92\\
         &                     & 50(30)& 0.04& 2     &-0.09&1.01&1.00& 0.02&0.86&0.99&-0.02&1.44&0.92\\
\hline
\hline
\end{tabular}
\label{table1}
\normalsize
\end{table}

\pagebreak\clearpage\newpage
\thispagestyle{empty}
\begin{table}[h]
\centering
\caption{Simulation study for RLRT of overall pathway effect with $\rho$ fixed at different values and estimated. Simulated samples size  $n=100$, and both used and true gene number equal to $p=30$.}
\begin{tabular}{ccccccc}
\hline\hline
                      &$\rho$    &$b=0$& 0.2    & 0.35 &  0.5& 1    \\
\hline
\multirow{4}{*}{$a=0$}& 2        & 0.03& 0.34   & 0.91 & 1.00& 1.00 \\
                      & 5        & 0.02& 0.34   & 0.89 & 0.99& 1.00 \\
                      & 10       & 0.02& 0.30   & 0.88 & 0.99& 1.00 \\
                      &estimated & 0.03& 0.33   & 0.87 & 0.99& 1.00 \\
\hline
                      &          &$a=0$& 0.05   & 0.1  & 0.2 & 0.5  \\\cline{2-7}
\multirow{4}{*}{$b=0$}& 2        & 0.03& 0.07   & 0.37 & 0.96& 1.00 \\
                      & 5        & 0.02& 0.07   & 0.37 & 0.95& 1.00 \\
                      & 10       & 0.02& 0.06   & 0.34 & 0.91& 1.00 \\
                      &estimated & 0.03& 0.06   & 0.34 & 0.93& 1.00 \\
\hline
\hline
\end{tabular}
\label{table2}
\end{table}

\pagebreak\clearpage\newpage
\thispagestyle{empty}
\begin{table}[h]
\centering
\caption{Simulation study for RLRT of overall pathway effect with fitted genes number $p$ equal or larger than true one $p=30$. Simulated samples size  $n=60$ and $n=35$. The parameter $\rho$ is fixed at 2.}
\begin{tabular}{cccccccc}
\hline\hline
                      &     $n$             &used $p$  &$b=0$& 0.2    & 0.35 &  0.5& 1     \\
\hline
\multirow{4}{*}{$a=0$}&\multirow{2}{*}{$60$}& 30       & 0.03& 0.18   & 0.57 & 0.88& 1.00 \\
                      &                     & 50       & 0.03& 0.15   & 0.48 & 0.76& 0.99 \\\cline{2-8}
                      &\multirow{2}{*}{$35$}& 30       & 0.04& 0.10   & 0.27 & 0.46& 0.85 \\
                      &                     & 50       & 0.03& 0.08   & 0.23 & 0.38& 0.78 \\
\hline
                      &                     &          &$a=0$& 0.1    & 0.2  & 0.5 & 1.5  \\\cline{3-8}
\multirow{4}{*}{$b=0$}&\multirow{2}{*}{$60$}& 30       & 0.03& 0.15   & 0.51 & 0.72& 0.72 \\
                      &                     & 50       & 0.03& 0.13   & 0.41 & 0.72& 0.76  \\\cline{2-8}
                      &\multirow{2}{*}{$35$}& 30       & 0.04& 0.09   & 0.25 & 0.56& 0.63 \\
                      &                     & 50       & 0.03& 0.05   & 0.18 & 0.43& 0.55  \\

\hline
\hline
\end{tabular}
\label{table3}
\end{table}

\pagebreak\clearpage\newpage
\thispagestyle{empty}
\begin{table}[h]
\centering
\caption{Simulation study for PLRT and score test of P-E interaction with $\rho$ fixed at different values. Fitted and used gene numbers are equal to $p=5$, and $n=100$.}
\begin{tabular}{ccccccccc}
\hline\hline
       &$\rho$    &$b=0$& 0.1    & 0.2  & 0.35&  0.5 & 0.8  & 1   \\
\hline
       & 2        & 0.04& 0.24   & 0.58 & 0.95& 1.00 & 1.00 & 1.00\\
RLRT   & 5        & 0.04& 0.24   & 0.64 & 0.98& 1.00 & 1.00 & 1.00\\
       & 10       & 0.03& 0.24   & 0.67 & 0.97& 1.00 & 1.00 & 1.00\\
\hline
score  & 2        & 0.08& 0.31   & 0.68 & 0.98& 1.00 & 1.00 & 1.00\\
test   & 5        & 0.06& 0.30   & 0.72 & 0.97& 1.00 & 1.00 & 1.00\\
       & 10       & 0.06& 0.26   & 0.72 & 0.98& 1.00 & 1.00 & 1.00\\
       \hline
\hline
\end{tabular}
\label{table4}
\end{table}

\pagebreak\clearpage\newpage
\thispagestyle{empty}
\begin{table}[h] 
\centering
\caption{Estimated parameters of top 20 pathways obtained from p-REML and ranked by $p$-values of testing RLRT $D$. The numbers in the round brackets are the standard errors.}
\scriptsize
\begin{tabular}{ccccccccccc}
\hline\hline
 pathway   &       &                 &                 &                  &              &              &                 & fixed  & RLRT  & RLRT       \\
  ID       & gene\#& $\hat{\beta}_0$ & $\hat{\beta}_1$ & $\hat{\sigma}^2$ & $\hat\tau_x$ & $\hat\tau_z$ & $\hat\tau_{xz}$ & $\rho$ & $D$   & $p$-value\\
\hline
  73       & 11  &5.09(1.51)& -0.01(0.21) &0.08(0.39) & 1.0e-11(0.02) & 6.09(3.12)     & 17.7(11.8)   & 0.457 & 12.2 & 0.001   \\
  274      & 16  &7.25(1.35)&  0.20(0.16) &0.66(0.71) & 2.1e-09(0.02) & 4.74(3.09)     & 9.74(8.90)   & 0.581 & 7.68 & 0.006   \\
  230      & 121 &5.69(1.39)&  0.15(0.14) &0.10(1.03) & 7.3e-11(0.02) & 5.75(3.42)     & 6.17(6.99)   & 0.330 & 7.81 & 0.006   \\
  229      & 133 &5.82(1.13)&  0.15(0.12) &1.29(1.28) & 1.7e-03(0.02) & 3.25(2.99)     & 3.96(6.31)   & 0.289 & 6.65 & 0.012   \\
  152      & 11  &6.13(1.12)&  0.21(0.15) &2.16(0.91) & 8.6e-09(0.02) & 1.57(2.21)     & 7.48(8.69)   & 1.266 & 6.20 & 0.014   \\
  16       & 49  &5.76(1.00)&  0.14(0.13) &1.98(1.24) & 1.5e-08(0.02) & 1.89(2.55)     & 4.57(6.82)   & 0.308 & 5.93 & 0.017   \\
  173      & 11  &6.06(1.07)&  0.19(0.15) &2.14(0.92) & 2.1e-09(0.01) & 1.57(2.22)     & 7.10(7.93)   & 0.756 & 5.77 & 0.017   \\
  236      & 22  &6.27(1.06)&  0.23(0.15) &2.10(1.06) & 1.4e-08(0.02) & 1.41(2.24)     & 7.24(8.08)   & 0.862 & 5.63 & 0.019   \\
  144      & 7   &5.43(1.21)&  0.15(0.20) &2.35(0.85) & 1.6e-03(0.02) & 1.16(2.26)     & 11.5(11.7)   & 0.411 & 5.35 & 0.019   \\
  151      & 20  &6.08(1.04)&  0.22(0.14) &2.15(1.06) & 7.5e-09(0.02) & 1.52(2.24)     & 6.21(7.51)   & 0.937 & 5.62 & 0.019   \\
  14       & 49  &6.09(1.20)&  0.16(0.14) &1.57(1.27) & 1.3e-09(0.02) & 2.76(2.91)     & 5.72(7.42)   & 0.706 & 5.30 & 0.024   \\
  228      & 43  &6.16(0.77)&  0.20(0.14) &2.88(1.18) & 7.4e-11(0.02) & 0.03(1.73)     & 5.91(6.86)   & 0.374 & 4.95 & 0.028   \\
  103      & 37  &6.09(0.90)&  0.20(0.14) &2.58(1.20) & 9.8e-09(0.02) & 0.74(2.08)     & 5.76(7.42)   & 0.751 & 4.82 & 0.030   \\
  271      & 37  &6.20(0.92)&  0.22(0.14) &2.45(1.23) & 7.5e-12(0.02) & 0.94(2.19)     & 5.73(7.19)   & 0.702 & 4.83 & 0.030   \\
  150      & 21  &5.98(0.94)&  0.19(0.14) &2.54(1.12) & 7.5e-11(0.02) & 0.97(2.10)     & 5.75(7.65)   & 1.161 & 4.66 & 0.033   \\
  172      & 8   &5.85(0.92)&  0.15(0.18) &2.75(0.99) & 2.6e-03(0.02) & 3.5e-10(1.61)  & 10.1(9.8)    & 0.812 & 4.22 & 0.039   \\
  133      & 58  &6.01(0.83)&  0.18(0.14) &2.71(1.29) & 1.8e-03(0.02) & 0.32(2.04)     & 6.28(7.18)   & 0.339 & 4.15 & 0.044   \\
  8        & 27  &5.87(0.78)&  0.18(0.15) &2.92(1.15) & 1.6e-02(0.04) & 3.0e-09(1.72)  & 5.96(7.21)   & 0.527 & 4.08 & 0.045   \\
  101      & 13  &6.08(0.90)&  0.19(0.16) &3.01(1.01) & 5.7e-10(0.02) & 0.23(1.59)     & 6.81(8.79)   & 0.458 & 3.88 & 0.045   \\
  158      & 8   &5.79(1.00)&  0.15(0.14) &2.55(0.98) & 1.3e-09(0.02) & 1.55(2.24)     & 5.39(7.72)   & 0.621 & 3.53 & 0.056   \\
\hline
\hline
\end{tabular}
\label{table5}
\normalsize
\end{table}

\pagebreak\clearpage\newpage
\thispagestyle{empty}
\begin{table}[h] 
\centering
\caption{P-values of different tests for top 20 pathway significant in the overall pathway effect. Columns 2 and 3 are labels indicating appearance in the top 50 list of other methods or not. Missing values in column 6 is because the information matrix is not positive definite.}
\begin{tabular}{ccccccccc}
\hline\hline
 pathway&Global       &Forest   & RLRT      & permutation  &RLRT        & permutation  &score test           \\
  ID    &Score Test   &Tree     & test for D& test for $D$ &test for $d$& test for $d$ &for $U_{\tau_{xz}}$  \\
\hline
  73   & Yes & Yes & 0.001     & 0.001    &0.002 &0.001   &   0.005  \\
  274  & Yes & No  & 0.006     & 0.011    &0.025 &0.013   &   0.016  \\
  230  & Yes & Yes & 0.006     & 0.010    &-     &0.025   &   0.007  \\
  229  & Yes & Yes & 0.012     & 0.020    &-     &0.138   &   0.062  \\
  152  & No  & No  & 0.014     & 0.015    &0.179 &0.303   &   0.163  \\
  16   & Yes & Yes & 0.017     & 0.027    &0.126 &0.147   &   0.058  \\
  173  & Yes & Yes & 0.017     & 0.020    &0.017 &0.018   &   0.002  \\
  236  & No  & No  & 0.019     & 0.021    &0.133 &0.119   &   0.104  \\
  144  & Yes & Yes & 0.019     & 0.020    &0.076 &0.072   &   0.106  \\
  151  & No  & No  & 0.019     & 0.023    &0.205 &0.262   &   0.146  \\
  14   & Yes & No  & 0.024     & 0.031    &0.113 &0.054   &   0.046  \\
  228  & Yes & Yes & 0.028     & 0.035    &0.032 &0.024   &   0.006  \\
  103  & No  & Yes & 0.030     & 0.039    &0.121 &0.106   &   0.086  \\
  271  & No  & No  & 0.030     & 0.037    &0.148 &0.142   &   0.110  \\
  150  & No  & No  & 0.033     & 0.034    &0.080 &0.062   &   0.044  \\
  172  & No  & No  & 0.039     & 0.044    &0.016 &0.015   &   0.009  \\
  133  & No  & No  & 0.044     & 0.057    &0.053 &0.043   &   0.018  \\
  8    & Yes & Yes & 0.045     & 0.052    &0.051 &0.038   &   0.032  \\
  101  & No  & No  & 0.045     & 0.044    &0.068 &0.049   &   0.056  \\
  158  & Yes & No  & 0.056     & 0.054    &-     &0.343   &   0.560  \\
\hline
\hline
\end{tabular}
\label{table6}
\end{table}

\pagebreak\clearpage\newpage
\thispagestyle{empty}

\pagebreak\clearpage\newpage
\thispagestyle{empty}

\pagebreak\clearpage\newpage
\thispagestyle{empty}

\pagebreak\clearpage\newpage
\thispagestyle{empty}

\pagebreak\clearpage\newpage
\thispagestyle{empty}
\Appendix
\section{The Representation of the Natural Cubic Spline}\label{ap1}
\markright{\slshape\small\hfill Appendix\hfill}
Following \citet{b9}, the representation of the natural cubic spline (\ref{a_e13}) in section \ref{a_sec2_3} is called the value-second derivative representation. Details for defining matrices $B$ and $M$ are shown as the following.

Suppose $f_x$ is the natural cubic spline with $n$ distinct $x^0_1<,...,<x^0_n$. Define
\[
f_{x,i}=f_x(x_i^0)\hbox{ and }\gamma_i=f_x''(x_i^0)\hbox{ for } i=1,...,n
\]
By the definition of natural cubic spline, $\gamma_1=\gamma_n=0$. Let $\mathbf f_x$ stands for the vector $(f_{x,1},...,f_{x,n})^T$ and let $\boldsymbol\gamma=(\gamma_2,...,\gamma_{n-1})^T$ where $\boldsymbol\gamma$ is a $(n-2)\times1$ vector with the element index starting at $i=2$. Now define two matrices, $Q$ and $R$. Let $h_i=t_{i+1}-t_i$ for $i=1,...,n-1$. Let $Q$ be the $n\times(n-2)$ matrix with entries $q_{ij}$, for $i=1,...,n-1$ and $j=2,...,n-1$, given by
\be
q_{j-1,j}=h_{j-1}^{-1},\;q_{jj}=-h_{j-1}^{-1}-h_j^{-1},\hbox{ and }q_{j+1,j}=h_j^{-1}
\ee
for $j=2,...,n-1$ and $q_{ij}=0$ for $|i-j|\ge2$. The columns of $Q$ are indexed in the same way as the elements of $\boldsymbol\gamma$ starting at $j=2$, so that the first element of $Q$ is $q_{12}$.

$R$ is a $(n-2)\times(n-2)$ symmetric matrix with elements $r_{ij}$, for $i$ and $j$ running from 2 to $n-1$, given by
\be
\begin{split}
r_{ii}={1\over3}(h_{i-1}+h_i)&\hbox{ for } i=2,...,n-1,\\
r_{i,i+1}=r_{i+1,i}={1\over6}h_i&\hbox{ for }i=2,...,n-2,
\end{split}
\ee
and $r_{ij}=0$ for $|i-j|\ge2$.

The matrix $R$ is strictly diagonal dominant and strictly positive definite. Using the Cholesky factorization that avoids taking the square roots \citep{b9} Section 2.6.1, we can factorize $R$ as
\[
R=U\Lambda U^T,
\]
where $\Lambda$ is a diagonal matrix and $U$ is a lower triangular band matrix with diagonal elements all equal to 1. Since $R$ are strictly positive definite, all diagonal elements of $\Lambda$ are positive, $R^{-1}=(\Lambda^{1/2}U^T)^{-1}(U\Lambda^{1/2})^{-1}$. The penalty matrix $M$ can be expressed as
\be
M=QR^{-1}Q^T=Q(\Lambda^{1/2}U^T)^{-1}(U\Lambda^{1/2})^{-1}Q^T=LL^T,
\ee
where $L=Q(\Lambda^{1/2}U^T)^{-1}$. The $B$ matrix thus is calculated by
\[
\begin{split}
B&=L(L^TL)^{-1}=Q(\Lambda^{1/2}U^T)^{-1}\left\{[(\Lambda^{1/2}U^T)^{-1}]^TQ^TQ(\Lambda^{1/2}U^T)^{-1}\right\}^{-1}\\
&=Q(\Lambda^{1/2}U^T)^{-1}(\Lambda^{1/2}U^T)(Q^TQ)^{-1}(\Lambda^{1/2}U^T)^T\\
&=Q(Q^TQ)^{-1}U\Lambda^{1/2}.
\end{split}
\]
The Theorem 2.1 in \citet{b9} states that the vectors $\mathbf f_x$ and $\boldsymbol\gamma$ specific a natural cubic spline $f_x$ if and only if the condition $Q^T\mathbf f_x=R\boldsymbol\gamma$ is satisfied. If this condition is satisfied then the roughness penalty will satisfy
\[
\begin{split}
\int_0^1\{f_x''(x)\}^2dx&=\sum_{j=1}^{n-1}{{\gamma_{j+1}-\gamma_j}\over{h_j}}(f_{x,j}-f_{x,j+1})=\boldsymbol{\gamma}^TQ^T\mathbf f_x\\
&=\boldsymbol\gamma^TR\boldsymbol\gamma=\mathbf f_x^TQR^{-1}Q^T\mathbf f_x=\mathbf f_xM\mathbf f_x.
\end{split}
\]
In the above derivation we assumed that $x^0_i, i=1,...,n$, were distinct and ordered, so the rank of the penalty matrix $M$ is $n-2$ and $B$ is a $n\times(n-2)$ matrix. In our model, we shall have $r$ distinct and ordered $x_i^0,i=1,...,r$, from the observed data $x_i, i=1,...,n$, where $r\le n$ and $x_i$'s may not be ordered. Based the $r$ $x_i^0$'s, $B$ is a $r\times(r-2)$ matrix. Thus we will use a $n\times r$ incidence matrix $N$ defined in a way similar to that given by \citet{b9}, Section 4.3.1, such that $B=NB$, where the left $B$ is what we shall use in the model, and the right $B$ is calculated based on $r$ distinct $x_i^0$'s.

\end{document}